\documentclass[a4paper,fleqn,usenatbib]{mnras}

\usepackage{newtxtext,newtxmath}

\usepackage[T1]{fontenc}
\usepackage{ae,aecompl}

\usepackage{graphicx}	
\usepackage{amsmath}	
\usepackage{amssymb}	

\newcommand\grad{{\bmath\nabla}}
\newcommand\bcdot{\cdot}
\newcommand\veczero{{\bmath0}}

\newcommand\vece{{\bmath e}}

\newcommand\vecg{{\bmath g}}

\newcommand\vecu{{\bmath u}}
\newcommand\vecv{{\bmath v}}

\newcommand\dd{\mathrm{d}}
\newcommand\DD{\mathrm{D}}
\newcommand\ee{\mathrm{e}}
\newcommand\ii{\mathrm{i}}

\newcommand\f{\frac}
\newcommand\p{\partial}
\newcommand\cst{\mathrm{constant}}

\newcommand\barvecx{\bar{\bmath x}}
\newcommand\gradbar{\bar{\bmath\nabla}}
\DeclareMathOperator{\sech}{sech}

\def\XXint#1#2#3{{\setbox0=\hbox{$#1{#2#3}{\int}$}
     \vcenter{\hbox{$#2#3$}}\kern-.5\wd0}}

\def\XXiint#1#2#3{{\setbox0=\hbox{$#1{#2#3}{\iint}$}
     \vcenter{\hbox{$#2#3$}}\kern-.5\wd0}}

\title[Gravitational instability of gaseous discs]
{Gravitational instability and affine dynamics of gaseous astrophysical discs}

\author[Gordon I. Ogilvie]
{Gordon I. Ogilvie\thanks{E-mail: gio10@cam.ac.uk}
\\
Department of Applied Mathematics and Theoretical Physics,
University of Cambridge, Centre for Mathematical Sciences,\\
Wilberforce Road, Cambridge CB3 0WA, UK
}

\date{Accepted 2025 March 13. Received 2025 March 10; in original form 2024 November 15}

\pubyear{2025}

\begin{document}
\label{firstpage}
\pagerange{\pageref{firstpage}--\pageref{lastpage}}
\maketitle

\begin{abstract}
We develop several aspects of the theory of gaseous astrophysical discs in which the gravity of the disc makes a significant contribution to its structure and dynamics. We show how the internal gravitational potential can be expanded in powers of the aspect ratio of the disc (or of a structure within it) and separated into near and far contributions. We analyse the hydrostatic vertical structure of a wide family of disc models, both analytically and numerically, and show that the near contribution to the internal gravitational potential energy can be written in an almost universal form in terms of the surface density and scaleheight. We thereby develop an affine model of the dynamics of (generally non-hydrostatic) self-gravitating discs in which this contribution to the energy acts as a gravitational pressure in the plane of the disc. This combines with and significantly reinforces the gas pressure, allowing us to define an enhanced effective sound speed and Toomre stability parameter $Q$ for self-gravitating discs. We confirm that this theory fairly accurately reproduces the onset of axisymmetric gravitational instability in discs with resolved vertical structure. Among other things, this analysis shows that the critical wavelength is on the order of twenty times the scaleheight, helping to justify the validity of the affine model. The weakly nonlinear theory also typically exhibits subcritical behaviour, with equilibrium solutions of finite amplitude being found in the linearly stable regime $Q>1$ for adiabatic exponents less than $1.50$.
\end{abstract}

\begin{keywords}
  accretion, accretion discs -- gravitation -- hydrodynamics -- instabilities -- waves -- methods: analytical
\end{keywords}



\section{Introduction}

Gaseous discs around young stars and in active galactic nuclei are often sufficiently massive that the self-gravity of the disc plays an important role in their structure and dynamics.
The main effect that is usually considered is gravitational instability \citep[GI, e.g.][]{2016ARA&A..54..271K,2003MNRAS.339..937G}, whereby an initially smooth disc that is sufficiently massive and cool develops structure, initially in the form of rings or spiral waves. The nonlinear outcome of GI is thought to be either a gravitational turbulence of trailing spiral density waves or a fragmentation of the disc into bound objects (protoplanets or protostars), depending on how the timescale on which the disc is able to cool compares with the orbital timescale \citep{2001ApJ...553..174G}. GI in gaseous discs is closely related to GI in discs of collisionless stars \citep[spiral galaxies;][]{1964ApJ...139.1217T} or collisional particles \citep[planetary rings;][] {2001Icar..154..296D}.

The classic analysis of GI considers a 2D gaseous disc, meaning that the structure in the vertical direction perpendicular to the plane of the disc is ignored or crudely `integrated'. This approach leads to the local dispersion relation
\begin{equation}
  \omega^2=\kappa^2-2\pi G\Sigma|k|+v_\text{s}^2k^2,
\end{equation}
relating the angular frequency $\omega$ of an axisymmetric density wave to its radial wavenumber $k$, in terms of the epicyclic frequency $\kappa$, the surface density $\Sigma$ and the sound speed $v_\text{s}$. This dispersion relation implies that the disc is unstable, for a band of wavelengths centred on $2\pi^2 G\Sigma/\kappa^2$, when the Toomre stability parameter
\begin{equation}
  Q=\f{\kappa v_\text{s}}{\pi G\Sigma}<1.
\end{equation}
While undoubtedly very useful as a rough estimate of the gravitational stability of a disc, $Q$ has important limitations, as many authors have previously recognized.

Numerical simulations of gravitational turbulence resulting from GI in 2D and 3D discs with a sufficiently long cooling timescale to avoid fragmentation, using either global or local computational models, show that there is a thermostatic regulation in the turbulent state in which $Q$ is of order unity, suggestive of marginal stability. However, it is debatable how this finding should be interpreted. First, it is not clear how best to define an appropriate (average) value of $Q$ in a turbulent state, given the variations of the sound speed and surface density, especially in a 3D disc, for which even the linear stability criterion is not well known. Second, it is not clear whether the axisymmetric stability criterion is directly relevant, given that non-axisymmetric density waves can undergo transient growth (known as swing amplification) when $Q>1$ \citep[e.g.][]{1981seng.proc..111T,1992MNRAS.256..685N}.
Indeed, one of the limitations of~$Q$ mentioned above is that the stability of non-axisymmetric modes can depend on an additional parameter \citep{1978ApJ...226..508L,2016ARA&A..54..271K}.

Since the foundational work of \citet{1965MNRAS.130...97G,1965MNRAS.130..125G}, relatively little detailed analytical work has been carried out on the 3D dynamics of self-gravitating gaseous discs. The 3D structure of self-gravitating discs has been studied by \citet{1999A&A...350..694B}, and axisymmetric linear modes of these discs have been computed by \citet{2010MNRAS.406.2050M}. Various techniques have been used to approximate the effects of non-zero thickness on the dynamics and stability of self-gravitating discs without an explicit  3D calculation; most commonly, a Plummer potential with a characteristic length related to the vertical scaleheight is introduced to soften the gravitational interaction between horizontally separated elements of the disc. Examples of fully 3D global and local simulations of GI using a variety of numerical methods are \citet{2009MNRAS.393.1157C}, \citet{2019MNRAS.483.3718B}, \citet{2021A&A...650A..49B}, \citet{2023ApJ...958..139S} and \citet{2023MNRAS.520.3097Z}.

We previously developed an affine model of the hydrodynamics of non-self-gravitating discs \citep{2018MNRAS.477.1744O}. In this approach, a thin disc is thought of as a family of columnar fluid elements, each of which can undergo a 3D translation as well as a 3D linear transformation (including expansion and shear), i.e.\ an affine transformation that depends on the column. This model captures the degrees of freedom involved in the large-scale dynamics of warped, eccentric and tidally distorted discs and its ideal gas dynamics can be derived from Hamilton's principle. A simplifying geometrical approximation is used that is valid when the scale of the deformation is large compared to the vertical scaleheight of the disc.

One of the aims of the present paper is to develop an affine model of the dynamics of self-gravitating discs in the case in which the disc remains symmetric about the midplane. (We defer a treatment of warped self-gravitating discs to future work.) Along the way, however, we develop some basic theory of the structure and stability of self-gravitating discs in 3D, which we could not find elsewhere in the literature. We first analyse the gravitational potential and energy of a thin 3D disc (Section~\ref{s:gravity}) and obtain new results on the vertical structure of discs in equilibrium (Section~\ref{s:equilibrium}). We then develop the affine model of dynamical self-gravitating discs (Section~\ref{s:affine}). We use the affine model, in conjunction with some exact results, to discuss the linear stability (Section~\ref{s:stability}) and weakly nonlinear dynamics (Section~\ref{s:nonlinear}) of self-gravitating discs, before summarizing and concluding (Section~\ref{s:conclusions}). Some more technical results are reported in the appendices.

\section{Gravity of a thin disc}
\label{s:gravity}

In this section we analyse the internal gravitational potential and potential energy of a thin disc. We develop an approximation for these quantities that can be regarded as an expansion in powers of the aspect ratio of the disc (or of the relevant structure within it) and goes beyond the usual 2D (`razor-thin') approximation. It also separates the near and far contributions to the self-gravity of the disc.

\subsection{Definition of the problem}

Consider a thin disc that has reflectional symmetry in the plane $z=0$, where $(x,y,z)$ are Cartesian coordinates. The disc experiences a total gravitational potential $\Phi=\Phi_\text{int}+\Phi_\text{ext}$ that is the sum of an internal potential generated by the disc itself, and an external potential due to the central body around which the disc orbits, as well as any other external masses.

The internal gravitational potential due to the disc is
\begin{equation}
  \Phi_\text{int}(x,y,z)=-G\iiint\f{\rho(x',y',z')}{\Delta_3}\,\dd x'\,\dd y'\,\dd z',
\label{phi_int_green}
\end{equation}
where $\rho$ is the density,
\begin{equation}
  \Delta_3=\sqrt{\left(x'-x\right)^2+\left(y'-y\right)^2+\left(z'-z\right)^2}
\end{equation}
is the 3D distance, and the integral extends over the entire disc.

Let us define the surface density $\Sigma(x,y)$ and the vertical scaleheight $H(x,y)$ of the disc by
\begin{equation}
  \Sigma=\int\rho\,\dd z,\qquad
  \Sigma H^2=\int\rho z^2\,\dd z,
\end{equation}
where the integrals are over the full vertical extent of the disc. This specific way of defining~$H$ has been found most useful for the dynamics of non-self-gravitating discs \citep[e.g.][]{2018MNRAS.477.1744O}.

The disc is thin when $H$ is small compared to the typical horizontal lengthscale $L$ on which the disc varies. In other words, the density varies more rapidly with~$z$ than with $x$ or~$y$. Let us introduce the small parameter $\epsilon\ll1$ that characterizes the typical ratio of $H$ to~$L$. (It will not be necessary to define $L$ or $\epsilon$ precisely.) In many applications $L$ can be identified with the radius $r$, but in some cases such as in the presence of a density wave, it may be less than~$r$.

\subsection{Expansion of the potential}
\label{s:expansion_potential}

In Appendix~\ref{s:appendix_potential} we argue that the internal gravitational potential, evaluated at a point within the disc where $z$ is (at most) comparable to $H$, can be expanded in powers of the aspect ratio~$\epsilon$ as
\begin{equation}
  \Phi_\text{int}=\Phi_1+\Phi_2+\Phi_3+O(\epsilon^4),
\label{expansion_potential}
\end{equation}
with
\begin{align}
  &\Phi_1(x,y)=-G\iint\f{\Sigma(x',y')}{\Delta_2}\,\dd x'\,\dd y',\label{phi1_integral}\\
  &\Phi_2(x,y,z)=+2\pi G\int\rho(x,y,z')\left|z'-z\right|\,\dd z',\label{phi2_integral}\\
  &\Phi_3(x,y,z)=+G\iint\f{\Sigma(x',y')\left[z^2+H(x',y')^2\right]}{2\Delta_2^3}\,\dd x'\,\dd y',\label{phi3_integral}
\end{align}
where
\begin{equation}
  \Delta_2=\sqrt{\left(x'-x\right)^2+\left(y'-y\right)^2}
\end{equation}
is the 2D (horizontal) distance and the double integrals are over the projection of the disc on the plane $z=0$.

Let us note several properties of this expansion. The leading term $\Phi_1$ (equation~\ref{phi1_integral}) is independent of $z$ and corresponds to the usual potential that is written down for a 2D (`razor-thin') disc, when evaluated in the plane of the disc. Although the integrand is singular at $(x',y')=(x,y)$ where $\Delta_2=0$, the singularity is integrable and does not require any regularization or smoothing.\footnote{This can be seen, for example, by rewriting the integral in polar coordinates $(r,\theta)$ centred on the point $(x,y)$, in which case the factor $1/\Delta_2=1/r$ is cancelled by the factor $r$ in the area element $r\,\dd r\,\dd\theta$.} Indeed, the potential $\Phi_1$ is dominated by contributions from distant material. It is the main contribution to the \textit{far self-gravity} of the disc.

The next term $\Phi_2$ (equation~\ref{phi2_integral}) depends only on the vertical distribution of mass at the horizontal location of the point at which the potential is measured. Indeed, it is the potential of a 1D mass distribution such as a stratified slab and is positive. It will play a key role throughout this paper, and we discuss it further in Section~\ref{s:gpe} below. We call it the \textit{near self-gravity} of the disc.

The last term $\Phi_3$ (equation~\ref{phi3_integral}) represents a small positive correction to~$\Phi_1$ due to the non-zero thickness of the disc, which on average increases the distance between mass elements and slightly diminishes the far self-gravity. The term $\Phi_3$ consists of two parts. The first part, involving $z^2$, produces a potential that depends on $z$ and therefore contributes to the vertical component of gravity within the disc. It can be thought of as a tidal potential. The second part, involving $H(x',y')^2$, is sensitive to the scaleheight of distant material as well as its surface density. In total, $\Phi_3$ represents the gravitational interaction between distant columns of the disc in which one column is treated as a monopole and the other as a quadrupole (whereas in $\Phi_1$ both as treated as monopoles). This results in the much steeper $\Delta_2^{-3}$ dependence of the integrand on distance. Indeed, the integral defining $\Phi_3$ is strongly singular at $(x',y')=(x,y)$ and requires regularization. It is possible to do this by considering the `Hadamard finite part' of this singular integral, which is explained in Appendix~\ref{s:appendix_potential} and extends the more familiar concept of the Cauchy principal value to a more strongly singular integral. In the next subsection, however, we will see a simpler way to determine $\Phi_3$.

We abbreviate the notation by writing $\Sigma'$ for $\Sigma(x',y')$, etc., and $\dd A$ for the area element $\dd x\,\dd y$. If we also define $I=\Sigma H^2$, then
\begin{align}
  \Phi_1&=-G\int\f{\Sigma'}{\Delta_2}\,\dd A',\\
  \Phi_3&=\Phi_{3\text{a}}+\Phi_{3\text{b}}=\f{1}{2}z^2\,G\int\f{\Sigma'}{\Delta_2^3}\,\dd A'+G\int\f{I'}{2\Delta_2^3}\,\dd A'.\label{phi3ab}
\end{align}
We combine these contributions to the far self-gravity into
\begin{equation}
  \Phi_\text{d}(x,y)+\f{1}{2}z^2\Psi_\text{d}(x,y),
\end{equation}
where $\Phi_\text{d}=\Phi_1+\Phi_{3\text{b}}$ is the potential in the midplane $z=0$ and
\begin{equation}
  \Psi_\text{d}=G\int\f{\Sigma'}{\Delta_2^3}\,\dd A'
\label{psi_d}
\end{equation}
is the coefficient of $\f{1}{2}z^2$, contributing to a tidal potential.

\subsection{Relation to Poisson's equation}

As well as the integral formula~(\ref{phi_int_green}), the potential is also related to the density via Poisson's equation
\begin{equation}
  \left(\f{\p^2}{\p x^2}+\f{\p^2}{\p y^2}+\f{\p^2}{\p z^2}\right)\Phi_\text{int}=4\pi G\rho.
\end{equation}
Bearing in mind the expansion~(\ref{expansion_potential}), the property that derivatives with respect to $x$ and $y$ are generally smaller by a factor of $O(\epsilon)$ than derivatives with respect to $z$, and the fact that $\Phi_1$ does not depend on $z$, we can see that the expression of Poisson's equation at zeroth order\footnote{Implicitly, we are adopting a scaling in which $\rho=O(\epsilon^0)$, which allows the density formally to be comparable with the Roche density and the Toomre parameter $Q$ to be of order unity.} in $\epsilon$ is
\begin{equation}
  \f{\p^2\Phi_2}{\p z^2}=4\pi G\rho,
\label{poisson0}
\end{equation}
and at first order,
\begin{equation}
  \left(\f{\p^2}{\p x^2}+\f{\p^2}{\p y^2}\right)\Phi_1+\f{\p^2\Phi_3}{\p z^2}=0.
\label{poisson1}
\end{equation}
Indeed, the expression~(\ref{phi2_integral}) for $\Phi_2$ can be regarded as the integral solution of the 1D Poisson equation~(\ref{poisson0}), local in the horizontal coordinates.

Thus
\begin{equation}
  \Psi_\text{d}=-\left(\f{\p^2}{\p x^2}+\f{\p^2}{\p y^2}\right)\Phi_1=-\nabla_\text{h}^2\Phi_1,
\label{psid_integral}
\end{equation}
where $\nabla_\text{h}^2$ is the horizontal Laplacian operator.
Therefore the singular integrals in equation~(\ref{phi3ab}) can be evaluated in terms of non-singular integrals via
\begin{align}
  &\int\f{\Sigma'}{\Delta_2^3}\,\dd A'=\nabla_\text{h}^2\int\f{\Sigma'}{\Delta_2}\,\dd A',\\
  &\int\f{I'}{2\Delta_2^3}\,\dd A'=\nabla_\text{h}^2\int\f{I'}{2\Delta_2}\,\dd A'.
\end{align}

\subsection{Expansion of the potential energy}
\label{s:expansion_gpe}

The internal gravitational potential energy of the disc is given in general by $\mathfrak{W}_\text{int}=\f{1}{2}\int\Phi_\text{int}\,\dd m$ \citep[e.g.][]{2016JPlPh..82c2001O}. Corresponding to the expansion~(\ref{expansion_potential}) of the potential, there is a similar expansion of the potential energy: $\mathfrak{W}_\text{int}=\mathfrak{W}_1+\mathfrak{W}_2+\mathfrak{W}_3+\cdots$,
with
\begin{align}
  &\mathfrak{W}_1=-G\iint\f{\Sigma\Sigma'}{2\Delta_2}\,\dd A\,\dd A',\label{w1_integral}\\
  &\mathfrak{W}_2=+\pi G\iiint\rho(x,y,z)\rho(x,y,z')\left|z'-z\right|\,\dd z\,\dd z'\,\dd A,\label{w2_integral}\\
  &\mathfrak{W}_3=+G\iint\f{\Sigma\Sigma'(H^2+H'^2)}{4\Delta_2^3}\,\dd A\,\dd A'.\label{w3_integral}
\end{align}
Owing to the symmetry of the integral in equation~(\ref{w3_integral}), $H^2$ and $H'^2$ can be used interchangeably; we choose a symmetrized form.

\subsection{Example: the Kuzmin disc}

A useful analytical example is provided by the Kuzmin disc \citep{1956K}.
The razor-thin version of the Kuzmin disc has the axisymmetric potential
\begin{equation}
  \Phi_\text{int}(r,z)=-\f{GM_\text{d}}{\sqrt{r^2+(a+|z|)^2}}
\label{phi_kuzmin}
\end{equation}
and surface density
\begin{equation}
  \Sigma(r)=\f{M_\text{d}a}{2\pi(r^2+a^2)^{3/2}},
\label{sigma_kuzmin}
\end{equation}
where $(r,\phi,z)$ are cylindrical polar coordinates. The disc has a total mass~$M_\text{d}$ and a characteristic radius~$a$.
Above the midplane $z=0$, $\Phi_\text{d}$ is simply the potential of a fictitious point mass~$M_\text{d}$ located on the axis of symmetry at a distance~$a$ below the midplane (and vice versa).

A resolved version of the Kuzmin disc \citep{1975PASJ...27..533M} has the potential
\begin{equation}
  \Phi_\text{int}(r,z)=-\f{GM_\text{d}}{\sqrt{r^2+(a+\sqrt{z^2+b^2})^2}},
\label{phi_kuzmin_resolved}
\end{equation}
where the parameter $b$ is the characteristic thickness of the disc. The corresponding density can be calculated from Poisson's equation
\begin{equation}
  \f{1}{r}\f{\p}{\p r}\left(r\f{\p\Phi_\text{int}}{\p r}\right)+\f{\p^2\Phi_\text{int}}{\p z^2}=4\pi G\rho
\end{equation}
and has a somewhat complicated analytical form, but is concentrated near the midplane if the disc is thin ($b\ll a$).

If we regard both $z$ and $b$ as quantities of order $\epsilon\ll1$ relative to $r$ and $a$, then the density can be expressed as
\begin{equation}
  \rho\approx\f{M_\text{d}ab^2}{4\pi(r^2+a^2)^{3/2}(z^2+b^2)^{3/2}},
\label{rho_kuzmin}
\end{equation}
with a fractional error of order $\epsilon^2$. The vertical integral of this approximate density does indeed give the surface density~(\ref{sigma_kuzmin}).

It is instructive to expand the potential~(\ref{phi_kuzmin_resolved}) similarly in powers of $\epsilon$ within the disc. We obtain\footnote{Formally regarding $M_\text{d}=O(\epsilon^1)$, consistent with $\rho=O(\epsilon^0)$.}
\begin{align}
  \Phi_\text{int}&=\Phi_1+\Phi_2+\Phi_3+\cdots\nonumber\\
  &=-\f{GM_\text{d}}{\sqrt{r^2+a^2}}+\f{GM_\text{d}a\sqrt{z^2+b^2}}{(r^2+a^2)^{3/2}}\nonumber\\
  &\qquad+\f{GM_\text{d}(r^2-2a^2)(z^2+b^2)}{2(r^2+a^2)^{5/2}}+\cdots,
\end{align}
which can be seen as equivalent to the expansion~(\ref{expansion_potential}).
Thus $\Phi_1$ is independent of $z$ and corresponds to the midplane potential~(\ref{phi_kuzmin}) of the razor-thin Kuzmin disc. It can be understood as a global potential related to the surface density through an integral over all radii (equation~\ref{phi1_integral}). Then $\Phi_2$ is a $z$-dependent potential that is related to the local (in $r$) vertical structure through Poisson's equation~(\ref{poisson0}), where $\rho$ is given by equation~(\ref{rho_kuzmin}). Note also that for $|z|\gg b$ (but still $|z|\ll r,a$), $\Phi_2\approx2\pi G\Sigma|z|$. Finally, $\Phi_3$ is a $z$-dependent potential which combines the tidal potential $\f{1}{2}z^2\Psi_\text{d}$ and a $z$-independent part. In terms of Poisson's equation, we have
\begin{equation}
  \f{1}{r}\f{\dd}{\dd r}\left(r\f{\dd\Phi_1}{\dd r}\right)+\f{\p^2\Phi_3}{\p z^2}=0,
\end{equation}
which is why
\begin{equation}
  \Psi_\text{d}=-\f{1}{r}\f{\dd}{\dd r}\left(r\f{\dd\Phi_1}{\dd r}\right)
\end{equation}
can be deduced from radial differentiation of $\Phi_1$. These equations are of course equivalent to (\ref{poisson1}) and (\ref{psid_integral}) in the axisymmetric case.

\subsection{Total gravitational potential}
\label{s:total}

The total gravitational potential is $\Phi_\text{tot}=\Phi_\text{int}+\Phi_\text{ext}$, where $\Phi_\text{ext}$ is the potential due to the central object and any other external masses. We consider here the case of a central potential $\Phi_\text{ext}(r,z)$ that is axisymmetric and also reflectionally symmetric in the midplane $z=0$. The simplest case is a spherical or point mass $M_\text{c}$, for which $\Phi_\text{ext}=-GM_\text{c}/\sqrt{r^2+z^2}$.

Within the thin disc, $\Phi_\text{ext}$ can be expanded in a Taylor series in $z$ to give
\begin{equation}
  \Phi_\text{ext}=\Phi_\text{c}(r)+\f{1}{2}z^2\Psi_\text{c}(r)+O(\epsilon^4),
\end{equation}
where
\begin{equation}
  \Psi_\text{c}=\f{\p^2\Phi_\text{c}}{\p z^2}\bigg|_{z=0}.
\end{equation}
The potential contributions from the central object and the far self-gravity of the disc can then be combined into
\begin{equation}
  \Phi(r)+\f{1}{2}z^2\Psi(r),
\end{equation}
with $\Phi=\Phi_\text{c}+\Phi_\text{d}$ and $\Psi=\Psi_\text{c}+\Psi_\text{d}$.

The angular velocity $\Omega(r)$ of a circular orbit of radius~$r$ in the midplane, taking into account the combined potentials of the central object and the disc, is given by
\begin{equation}
  r\Omega^2=\f{\dd\Phi}{\dd r}.
\end{equation}
The angular frequencies $\kappa(r)$ and $\nu(r)$ of small horizontal and vertical perturbations, respectively, of the circular orbit are given by
\begin{align}
  &\kappa^2=\f{1}{r^3}\f{\dd(r^4\Omega^2)}{\dd r}=2\Omega(2\Omega-S),\\
  &\nu^2=\Psi,
\end{align}
where
$S=-r\,\dd\Omega/\dd r$
is the orbital shear rate.

For example, if a central point mass $M_\text{c}$ is combined with a Kuzmin disc of mass $M_\text{d}$, the midplane potential is
\begin{equation}
  \Phi=-\f{GM_\text{c}}{r}-\f{GM_\text{d}}{\sqrt{r^2+a^2}}.
\end{equation}
We also have
\begin{align}
  &\Omega^2=\f{GM_\text{c}}{r^3}+\f{GM_\text{d}}{(r^2+a^2)^{3/2}},\\
  &\kappa^2=\f{GM_\text{c}}{r^3}+\f{GM_\text{d}(r^2+4a^2)}{(r^2+a^2)^{5/2}},\\
  &\nu^2=\f{GM_\text{c}}{r^3}+\f{GM_\text{d}(r^2-2a^2)}{(r^2+a^2)^{5/2}}.
\end{align}
These satisfy the relation $2\Omega^2=\kappa^2+\nu^2$; this is because the Kuzmin potential~(\ref{phi_kuzmin}) satisfies Laplace's equation in the limit $z\to0$. Note also that $\nu^2<\Omega^2<\kappa^2$, so apsidal precession is retrograde and nodal precession is prograde. The ratio $\kappa^2/\nu^2$ has a peak value (at $r\approx1.22a$) of approximately $1+1.12(M_\text{d}/M_\text{c})$, if the mass ratio $M_\text{d}/M_\text{c}$ is reasonably small.

It is therefore reasonable to assume, for self-gravitating discs in which $M_\text{d}/M_\text{c}$ is fairly small, that $\kappa^2/\nu^2$ is close to unity, with a departure of order $M_\text{d}/M_\text{c}$.

\section{Equilibrium vertical structure}
\label{s:equilibrium}

Here we consider the relations between the density, pressure and gravitational potential that define the vertical structure of a self-gravitating disc in an equilibrium state. We work in the local approximation of astrophysical discs, also known as the shearing sheet or shearing box. This is appropriate because the thin disc is structured vertically on a length-scale that is small compared to the radius.

\subsection{Hydrostatic equilibrium and virial relation}

We therefore consider the dynamics of the disc in the neighbourhood of a circular reference orbit, where the local angular velocity, epicyclic frequency and vertical oscillation frequency associated with the central potential and the disc as a whole are $\Omega$, $\kappa$ and $\nu$, respectively. A local rotating Cartesian coordinate system is used, with $x$, $y$ and $z$ being measured in the radial, azimuthal and vertical directions with respect to the orbiting reference point.

An equilibrium state in the local model consists of a horizontally uniform disc with a hydrostatic vertical structure for the density $\rho(z)$, pressure $p(z)$ and near self-gravitational potential\footnote{The symbol $\Phi$ used in this section corresponds to $\Phi_2$ in Section~\ref{s:gravity}. Note that $\Phi_1$ and $\Phi_3$ contribute to the definitions of $\Omega$, $\kappa$ and $\nu$.} $\Phi(z)$. The equilibrium structure satisfies hydrostatic balance,
\begin{equation}
  -\f{1}{\rho}\f{\dd p}{\dd z}=\f{\dd\Phi}{\dd z}+\nu^2z,
\label{hydrostatic}
\end{equation}
and Poisson's equation,
\begin{equation}
  \f{\dd^2\Phi}{\dd z^2}=4\pi G\rho,
\label{poisson}
\end{equation}
and has reflectional symmetry in the midplane $z=0$. A further constraint is needed to connect the pressure to the density; this could be taken to be a polytropic or isothermal relation, as we will consider further below and in Appendix~\ref{s:appendix_polytropic}.

The appropriate solution of Poisson's equation~(\ref{poisson}) in 1D by means of Green's function is
\begin{equation}
  \Phi(z)=2\pi G\int\left|z-z'\right|\,\rho(z')\,\dd z'
=2\pi G\int\left|z-z'\right|\,\dd\mu',
\end{equation}
where the integral is over the full vertical extent of the disc and $\dd\mu=\rho\,\dd z$ is an element of mass per unit area. (This integral is equivalent to equation~\ref{phi2_integral} in Section~\ref{s:gravity}.) The corresponding vertical self-gravitational acceleration, apart from a minus sign, is
\begin{equation}
  \f{\dd\Phi}{\dd z}=2\pi G\int\text{sgn}\left(z-z'\right)\,\dd\mu'
\end{equation}
and the corresponding potential energy of the disc, per unit area, is
\begin{equation}
  W=\f{1}{2}\int\Phi\,\dd\mu=\pi G\iint\left|z-z'\right|\,\dd\mu\,\dd\mu'.
\label{w}
\end{equation}
The fact that $W$ is positive is a peculiarity of gravity in 1D. (Note that $W$ is the amount per unit area of the potential energy contribution $\mathfrak{W}_2$ given by equation~\ref{w2_integral} in Section~\ref{s:gravity}.)

Let us define the surface density $\Sigma$, scaleheight $H$ (as in Section~\ref{s:gravity}) and vertically integrated pressure $P$ by
\begin{equation}
  \Sigma=\int\dd\mu,\qquad
  \Sigma H^2=\int z^2\,\dd\mu,\qquad
  P=\int p\,\dd z.
\label{sigma_h_p}
\end{equation}
A virial relation can be obtained by multiplying the hydrostatic equation~(\ref{hydrostatic}) by $z$ and integrating with respect to~$\mu$. 
The pressure term can be integrated by parts, using the boundary condition that $zp\to0$ as $|z|\to\infty$:
\begin{equation}
  -\int z\f{\dd p}{\dd z}\,\dd z=\int p\,\dd z=P.
\end{equation}
The self-gravitational term can be manipulated as follows:
\begin{align}
  \int z\f{\dd\Phi}{\dd z}\,\dd\mu&=2\pi G\iint z\,\text{sgn}\left(z-z'\right)\,\dd\mu\,\dd\mu'\nonumber\\
  &=\pi G\iint\left(z-z'\right)\,\text{sgn}\left(z-z'\right)\,\dd\mu\,\dd\mu'\nonumber\\
  &=\pi G\iint\left|z-z'\right|\,\dd\mu\,\dd\mu'\nonumber\\
  &=W,
\end{align}
where, in the second line, we have symmetrized the double integrand, replacing $2f(z,z')$ with $f(z,z')+f(z',z)$. Hence we obtain the virial relation
\begin{equation}
  P=W+\nu^2\Sigma H^2,
\label{virial}
\end{equation}
which implies that pressure (which would cause the disc to expand vertically) is opposed by the sum of external gravity and self-gravity (both near and far), which would cause it to contract.

Let us define the dimensionless parameter
$s=W/P$,
which is the relative contribution of near self-gravity to the hydrostatic balance in the virial relation~(\ref{virial}). Then $P$ is partitioned according to
\begin{equation}
  W=sP,\qquad
  \nu^2\Sigma H^2=(1-s)P,
\label{s}
\end{equation}
with $0<s<1$. The limit $s\to0$ corresponds to a \textit{non-self-gravitating (NSG) disc} while the limit $s\to1$ corresponds to a \textit{purely self-gravitating (PSG) disc}.\footnote{Strictly speaking, $s$ refers to the relative importance of near self-gravity only. The relative contributions of external gravity and far self-gravity to $\nu^2=\Psi$ can be seen in the decomposition $\Psi=\Psi_\text{c}+\Psi_\text{d}$ discussed in Section~\ref{s:total}.}

\subsection{Dimensionless vertical structure}

Let us introduce the dimensionless vertical coordinate
$\zeta=z/H$
and write the density as
\begin{equation}
  \rho=\f{\Sigma}{H}F_\rho(\zeta),
\end{equation}
where $F_\rho(\zeta)$ is a dimensionless density profile. The mass element is then
\begin{equation}
  \dd\mu=\rho\,\dd z=\Sigma\,F_\rho\,\dd\zeta.
\end{equation}
For compatibility with the definitions in equation~(\ref{sigma_h_p}), the function $F_\rho(\zeta)$ must satisfy the normalization conditions
\begin{equation}
  \int F_\rho\,\dd\zeta=\int\zeta^2F_\rho\,\dd\zeta=1.
\label{normalization}
\end{equation}

Specific examples of density structure functions $F_\rho(\zeta)$ are discussed in Appendix~\ref{s:appendix_polytropic}, where we collect a large number of analytical and numerical results. In order to determine $F_\rho(\zeta)$ we need to make an assumption about the thermal structure of the disc, so that the pressure can be related to the density. A useful set of models, analogous to those used in stellar structure, are the polytropic discs for which
\begin{equation}
  p=K_3\rho^{1+1/n},
\label{polytrope_3d}
\end{equation}
where $n>0$ is the polytropic index (not necessarily an integer) and $K_3>0$ is the (3D) polytropic constant. A vertically polytropic disc is one in which $K_3$ is independent of $z$ for a particular choice of $n$.

In the limit $n\to0$, the polytropic disc becomes a homogeneous disc in which the density is independent of $z$, although the pressure does vary with $z$.

In the limit $n\to\infty$, the polytropic disc becomes an isothermal disc in which the temperature ($\propto p/\rho$ for a perfect gas) is independent of $z$.

\begin{figure}
  \includegraphics[width=0.9\columnwidth]{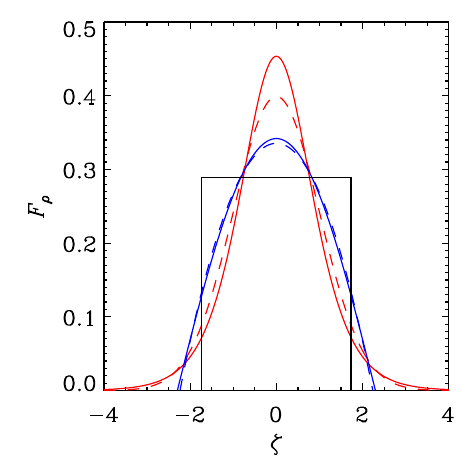}
  \caption{Examples of density structure functions giving the normalized vertical density profiles of discs. Black line: homogeneous disc (equation~\ref{density_homogeneous}). Red lines: isothermal disc in PSG (solid; equation~\ref{density_isothermal_psg}) and NSG (dashed; equation~\ref{density_isothermal_nsg}) limits. Blue lines: $n=1$ polytrope in PSG (solid; equation~\ref{density_n=1_psg}) and NSG (dashed; equation~\ref{density_n=1_nsg}) limits.}
\label{f:vertical_structure}
\end{figure}

As described in Appendix~\ref{s:appendix_polytropic}, the density structure function $F_\rho(\zeta)$ of polytropic models depends on both the polytropic index~$n$ and the degree of self-gravity~$s$. Some examples are plotted in Fig.~\ref{f:vertical_structure}. It can be seen that the homogeneous and isothermal models are two extremes that bracket a wide range of behaviour.

\subsection{Gravitational potential energy}
\label{s:gpe}

Having introduced $F_\rho(\zeta)$, we can write the near self-gravitational potential energy~(\ref{w}) of the equilibrium state, per unit area, as
\begin{equation}
  W=\mathcal{W}\pi G\Sigma^2H,
\label{w_c}
\end{equation}
where
\begin{equation}
  \mathcal{W}=\iint\left|\zeta-\zeta'\right|F_\rho(\zeta)F_\rho(\zeta')\,\dd\zeta\,\dd\zeta'
\end{equation}
is a positive dimensionless number of order unity.

The value of $\mathcal{W}$ varies so little between different models of the vertical structure that it could quite well be approximated as a universal constant. To illustrate this remarkable fact, we consider the polytropic models mentioned above and described in detail in Appendix~\ref{s:appendix_polytropic}. The value of $\mathcal{W}$ in such models, $\mathcal{W}_n(s)$, depends on the polytropic index, $0<n<\infty$, and the degree of self-gravity, $0<s<1$.

In the NSG limit $s\to0$, the value of $\mathcal{W}$ can be determined analytically (equation~\ref{c_nsg}).
$\mathcal{W}_n(0)$ is a monotonically decreasing function of $n$, the two extremes being
\begin{equation}
  \mathcal{W}_0(0)=\f{2}{\sqrt{3}}\approx1.1547,\qquad
  \mathcal{W}_\infty(0)=\f{2}{\sqrt{\pi}}\approx1.1284,
\end{equation}
for homogeneous and isothermal discs, respectively.

In the PSG limit $s\to1$, the value of $\mathcal{W}$ can also be determined analytically (equation~\ref{c_psg}).
$\mathcal{W}_n(0)$ is a monotonically decreasing function of $n$, the two extremes being
\begin{equation}
  \mathcal{W}_0(1)=\f{2}{\sqrt{3}}\approx1.1547,\qquad
  \mathcal{W}_\infty(1)=\f{2\sqrt{3}}{\pi}\approx1.1027,
\end{equation}
for homogeneous and isothermal discs, respectively.

In fact, the homogeneous disc has the same uniform density profile,
and the same $\mathcal{W}_0(s)=2/\sqrt{3}\approx1.1547$,
for any degree of self-gravity.

\begin{figure}
\includegraphics[width=0.9\columnwidth]{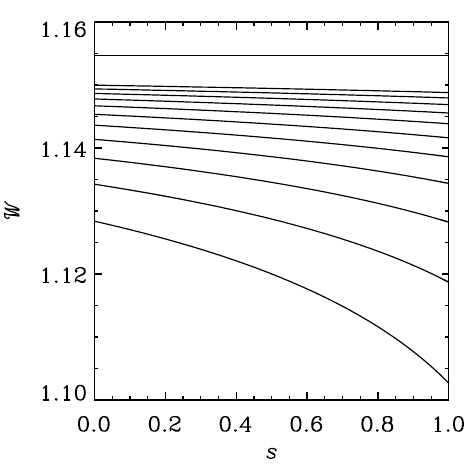}
\caption{Variation of the dimensionless gravitational energy $\mathcal{W}$ with the degree of self-gravity $s$ for polytropic equilibria with  (from bottom to top) $\gamma=1,1.1,1.2,1.3,1.4,1.5,1.6,1.7,1.8,1.9,2,\infty$, corresponding to $n=\infty,10,5,10/3,5/2,2,5/3,10/7,5/4,10/9,1,0$, respectively. The adiabatic exponent~$\gamma$ is related to the polytropic index~$n$ by equation~(\ref{gamma_n}).}
\label{f:w}
\end{figure}

Numerical integrations of the vertical structure of polytropic models (see Appendix~\ref{s:appendix_polytropic}) show that the variation of $\mathcal{W}_n(s)$ with $s$ increases with $n$, and is greatest for isothermal discs (Fig.~\ref{f:w}). Three further analytical reference points are useful. The smallest value of $\mathcal{W}$ is found for the purely self-gravitating isothermal disc [with the well-known $\sech^2$ density profile \citep{1942ApJ....95..329S}]:
\begin{equation}
  \mathcal{W}_\infty(1)=\f{2\sqrt{3}}{\pi}\approx1.1027,
\end{equation}
which differs from $\mathcal{W}_\infty(0)$ by about $2.3\%$. For the $n=1$ polytrope, which has an analytical solution for any $s$, the extremes are
\begin{equation}
  \mathcal{W}_1(0)=\f{18}{7\sqrt{5}}\approx1.1500,\qquad
  \mathcal{W}_1(1)=\f{\pi}{2\sqrt{\pi^2-8}}\approx1.1488,
\end{equation}
which differ by just over $0.1\%$.

In summary, the most extreme values of $\mathcal{W}$, corresponding to the homogeneous disc and the purely-self-gravitating isothermal disc, differ by less than $5\%$ and, for more realistic structures, the variation is much smaller than this. A universal approximation such as
$\mathcal{W}\approx1.15$
is likely to be very accurate.

\subsection{Entropy and 2D polytropic coefficient}

In a polytropic model the pressure $p$ and density $\rho$ are related by the power law~(\ref{polytrope_3d}). Consider a sequence of polytropic models in which $K_3$ and $n$ are fixed but $s$ varies.
How are the vertically integrated pressure $P$ and surface density $\Sigma$ related along this sequence? Let us define a 2D polytropic coefficient $K_2$ by
\begin{equation}
  \f{P}{H}=K_2\left(\f{\Sigma}{H}\right)^{1+1/n},
\label{polytrope_2d}
\end{equation}
which is analogous to equation~(\ref{polytrope_3d}) but uses vertically integrated quantities. In general there is no reason for $K_2$ to be identical to $K_3$, or even for it to remain constant along the sequence. As defined here, $K_2$ is a function of $s$ (as well as of $n$).

However, a second remarkable property, also discussed in detail in Appendix~\ref{s:appendix_polytropic}, is that $K_2$ is very nearly independent of~$s$, for a given value of~$n$. We express this property in terms of the dimensionless entropy\footnote{This name is appropriate because, if the polytrope is isentropic so that $n=1/(\gamma-1)$, $\mathcal{S}$ is equal to $p^n/\rho^{n+1}$ in units defined by $P$, $\Sigma$ and $H$, and this is proportional to the exponential of the entropy per particle in units of Boltzmann's constant.}
\begin{equation}
  \mathcal{S}=\left(\f{K_3}{K_2}\right)^n
\end{equation}
of the polytropic models.

In the NSG limit $s\to0$, the value of $\mathcal{S}$ can be determined analytically (equation~\ref{fs_nsg}). $\mathcal{S}_n(0)$ is a monotonically increasing function of $n$, the two extremes being
\begin{equation}
  \mathcal{S}_0(0)=2\sqrt{3}=3.4641,\qquad
  \mathcal{S}_\infty(0)=\sqrt{2\ee\pi}=4.1327,
\end{equation}
a variation of less than $20\%$.

In the PSG limit $s\to1$, the value of $\mathcal{S}$ can also be determined analytically (equation~\ref{fs_psg}). $\mathcal{S}_n(1)$ is a monotonically increasing function of $n$, the two extremes being
\begin{equation}
  \mathcal{S}_0(1)=2\sqrt{3}=3.4641,\qquad
  \mathcal{S}_\infty(1)=\f{\sqrt{3}\ee^2}{\pi}=4.0738,
\end{equation}
a variation of less than $20\%$.

\begin{figure}
\includegraphics[width=0.9\columnwidth]{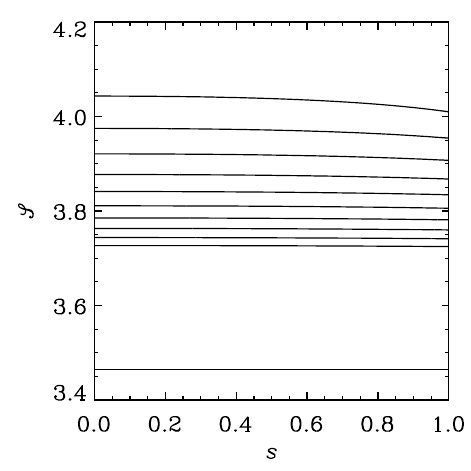}
\caption{Variation of the dimensionless entropy $\mathcal{S}$ with the degree of self-gravity $s$ for polytropic equilibria with (from top to bottom) $\gamma=1.1,1.2,1.3,1.4,1.5,1.6,1.7,1.8,1.9,2,\infty$.}
\label{f:s}
\end{figure}

For fixed $n>0$, $\mathcal{S}$ decreases monotonically with $s$, but the variation is much smaller than that with $n$ (Fig.~\ref{f:s}). For example, in the case $n=1$, $\mathcal{S}$ decreases monotonically from $5\sqrt{5}/3=3.7268$ at $s=0$ to $16/\pi\sqrt{\pi^2-8}=3.7247$ at $s=1$, a variation of about $0.05\%$. The difference between NSG and PSG values of $\mathcal{S}$ for a given $n$ is largest in the case of an isothermal disc, in which case it is less than $1.5\%$.

\subsection{An isentropic equilibrium sequence}
\label{s:isentropic}

Consider a family of vertically polytropic discs having the same values of the polytropic constant ($K_3$) and index ($n$), and experiencing the same external gravity and far self-gravity ($\nu^2$), but having differing degrees of near self-gravity ($s$) because of their different surface density ($\Sigma$).

We have seen that the algebraic relations
\begin{align}
  &P=\mathcal{W}\pi G\Sigma^2H+\nu^2\Sigma H^2,\label{family1}\\
  &\f{P}{H}=K_2\left(\f{\Sigma}{H}\right)^{1+1/n},\label{family2}
\end{align}
follow from hydrostatic equilibrium and the polytropic condition. Furthermore, we have seen that the values of $\mathcal{W}$ and $\mathcal{S}$ hardly vary with $s$. We may therefore treat $\mathcal{W}$ and $K_2$ as constants within the family of solutions, to a high degree of accuracy.

If we consider discs in which the specific entropy is independent of $z$, then the exponent of the polytropic power-law relation~(\ref{polytrope_3d}) can be identified with the adiabatic exponent of the gas:
\begin{equation}
  \gamma=1+\f{1}{n}.
\label{gamma_n}
\end{equation}
Indeed, this provides a good physical reason for considering a family of equilibria with the same $K_3$: one can be accessed from another by an adiabatic process in which each fluid element preserves its specific entropy.

An explicit solution of equations (\ref{family1})--(\ref{family2}) and the definition~(\ref{s}) of~$s$ is
\begin{align}
  &H^2=\f{K_2}{\nu^2}\left(\f{\nu^2}{\mathcal{W}\pi G}\right)^{\gamma-1}s^{\gamma-1}(1-s)^{2-\gamma},\label{h_s}\\
  &\Sigma^2=\f{K_2}{\nu^2}\left(\f{\nu^2}{\mathcal{W}\pi G}\right)^{\gamma+1}s^{\gamma+1}(1-s)^{-\gamma},\label{sigma_s}\\
  &P^2=\f{K_2^3}{\nu^2}\left(\f{\nu^2}{\mathcal{W}\pi G}\right)^{3\gamma-1}s^{3\gamma-1}(1-s)^{2-3\gamma}.\label{p_s}
\end{align}
We will see in Section~\ref{s:affine} below that the near self-gravitational energy~$W$ gives rise to a (2D) gravitational pressure, also equal to $W$, acting in the plane of the disc. We define the (vertically integrated) \emph{total pressure}
\[
  \Pi=P+W=(1+s)P,
\]
which is then given by
\begin{equation}
  \Pi^2=\f{K_2^3}{\nu^2}\left(\f{\nu^2}{\mathcal{W}\pi G}\right)^{3\gamma-1}s^{3\gamma-1}(1-s)^{2-3\gamma}(1+s)^2.\label{pi_s}
\end{equation}
These relations imply that $\Sigma$, $P$ and $\Pi$ are monotonically increasing functions of $s$, ranging from $0$ to $\infty$ as $s$ ranges from $0$ to $1$. (The behaviour of $H$ with $s$ depends on the value of $\gamma$.)

The variable $s$ provides a convenient dimensionless parametrization of the family of equilibria. Its logarithmic derivative with respect to  the surface density is
\begin{equation}
  \f{\dd\ln s}{\dd\ln\Sigma}=\f{2(1-s)}{\gamma+1-s}
\end{equation}
and is positive for $0<s<1$, implying that $s$ increases monotonically with $\Sigma$ (and vice versa).

Differentiating equations (\ref{h_s})--(\ref{p_s}), we find
\begin{equation}
  \f{\dd\ln H}{\dd\ln\Sigma}=\f{\gamma-1-s}{\gamma+1-s},\qquad
  \f{\dd\ln P}{\dd\ln\Sigma}=\f{3\gamma-1-s}{\gamma+1-s}.
\label{logderivs}
\end{equation}
The latter is positive, increasing monotonically from $(3\gamma-1)/(\gamma+1)$ as $s\to0$ to $(3\gamma-2)/\gamma$ as $s\to1$, except in the case $\gamma=1$, where it is constant and also equal to~$1$. These limiting values of $\dd\ln P/\dd\ln\Sigma$ for NSG and PSG discs are mentioned in \citet{2001ApJ...553..174G}.

The relation of the total pressure $\Pi$ to the surface density $\Sigma$, within the isentropic sequence, differs from a pure power law. We can define the effective 2D adiabatic exponent as the logarithmic derivative
\begin{equation}
  \Gamma^{(1)}=\f{\dd\ln\Pi}{\dd\ln\Sigma}=\f{3\gamma-1+3\gamma s-3s^2}{(1+s)(\gamma+1-s)}.
\end{equation}
This is also positive, ranging (but not generally monotonically) from $(3\gamma-1)/(\gamma+1)$ as $s\to0$ to $(3\gamma-2)/\gamma$ as $s\to1$. A measure of the departure of the dependence of $\Pi$ on $\Sigma$ from a pure power law is provided by the higher logarithmic derivatives
\begin{equation}
  \Gamma^{(j)}=\f{\dd^j\ln\Pi}{\dd(\ln\Sigma)^j}.
\label{higher_log}
\end{equation}
Typical values of $\Gamma^{(2)}$ and $\Gamma^{(3)}$ are of order $0.1$ or smaller.

It will be useful for what follows to define the \textit{effective sound speed} $c$ of the disc by
\begin{equation}
  c^2=\f{\dd\Pi}{\dd\Sigma}.
\end{equation}
For the isentropic sequence, this evaluates to
\begin{align}
  \label{csq_isentropic}
  c^2&=\f{(3\gamma-1+3\gamma s-3s^2)}{(1-s)(\gamma+1-s)}\nu^2H^2\\
  &=\f{(3\gamma-1+3\gamma s-3s^2)}{s(\gamma+1-s)}\mathcal{W}\pi G\Sigma H.
\end{align}
The first and second expressions for $c^2$ remain valid in the NSG and PSG limits, respectively.

The energy per unit mass of the disc,
\begin{equation}
  E=\f{P}{(\gamma-1)\Sigma}+\f{W}{\Sigma}+\f{1}{2}\nu^2H^2,
\end{equation}
evaluates to
\begin{equation}
  \left[\f{1}{(\gamma-1)}+\f{1}{2}(1+s)\right]K_2\left(\f{\nu^2}{\mathcal{W}\pi G}\right)^{\gamma-1}\left(\f{s}{1-s}\right)^{\gamma-1}.
\end{equation}
It is not difficult to show the differential relation
$\dd E=-\Pi\,\dd A$,
where $A=\Sigma^{-1}$ is the specific area and the differentials are taken along the isentropic sequence. Then we can define the specific enthalpy $\Upsilon=E+\Pi A$ such that $\dd\Upsilon=A\,\dd\Pi$ or, equivalently, $\dd\Pi=\Sigma\,\dd\Upsilon$. The specific enthalpy $\Upsilon$ evaluates to
\begin{equation}
  \left[\f{1}{(\gamma-1)}+\f{3}{2}(1+s)\right]K_2\left(\f{\nu^2}{\mathcal{W}\pi G}\right)^{\gamma-1}\left(\f{s}{1-s}\right)^{\gamma-1}.
\end{equation}

\subsection{The incompressible limit}

As $\gamma\to\infty$ (or $n\to0$), the derivatives $\dd\ln H/\dd\ln\Sigma$ and $\dd\ln P/\dd\ln\Sigma$ approach $1$ and $3$, respectively, independent of~$s$. In this limit we obtain a family of equilibria of equal (and uniform) density $\rho$, for which indeed $H\propto\Sigma$ and $P\propto\Sigma^3$. The exact relations are (see Appendix~\ref{s:appendix_homogeneous})
\begin{align}
  &H=\f{\Sigma}{2\sqrt{3}\,\rho},\\
  &P=\f{\pi G\Sigma^3}{3\rho}\left(1+\f{\nu^2}{4\pi G\rho}\right),\\
  &\Pi=\f{\pi G\Sigma^3}{3\rho}\left(2+\f{\nu^2}{4\pi G\rho}\right).
\end{align}
The specific energy and enthalpy are given by $E=\Pi/2\Sigma$ and $\Upsilon=3E$. Even though the fluid is incompressible in 3D, it retains a compressibility in 2D, with both $\dd\ln P/\dd\ln\Sigma$ and $\dd\ln\Pi/\dd\ln\Sigma$ being equal to $3$, and both density waves and GI are possible.

However, the parameter $s$ cannot be used to label the members of the family, because it is constant within the family:
\begin{equation}
  s=\left(1+\f{\nu^2}{4\pi G\rho}\right)^{-1}.
\end{equation}
In the case of a point-mass central potential with $\nu=\Omega$, the classical Roche limit, at which an incompressible fluid satellite is marginally disrupted by tidal forces, corresponds to $4\pi G\rho/\nu^2\approx44.4$. For a continuous disc, self-gravity can be important at densities several times smaller than this (see Appendix~\ref{s:appendix_homogeneous}).

\subsection{An isothermal sequence}
\label{s:isothermal}

In the limit $\gamma\to1$ (or $n\to\infty$), we obtain a family of isothermal equilibria of equal (and uniform) isothermal sound speed $c_\text{s}=\sqrt{p/\rho}$. This family is physically relevant for discs that undergo thermal relaxation towards an externally imposed temperature. The equilibrium is then defined by
\begin{equation}
  P=c_\text{s}^2\Sigma=\mathcal{W}\pi G\Sigma^2H+\nu^2\Sigma H^2,
\label{quadratic_isothermal}
\end{equation}
which is a quadratic equation for $H$
with one positive solution,\footnote{Equation~(\ref{h_isothermal}) has a similar mathematical form to equation~11 of \citet{2016ARA&A..54..271K}, based on the interpolation formula of \citet{1999A&A...350..694B}. The  coefficients differ because our treatment is more precise and we have used a particular definition of the scaleheight.}
\begin{equation}
  H=\left[\left(\f{\mathcal{W}\pi G\Sigma}{2\nu^2}\right)^2+\f{c_\text{s}^2}{\nu^2}\right]^{1/2}-\f{\mathcal{W}\pi G\Sigma}{2\nu^2}.
\label{h_isothermal}
\end{equation}
The degree of self-gravity is
\begin{equation}
  s=\f{W}{P}=\f{\mathcal{W}\pi G\Sigma}{c_\text{s}^2}H,
\end{equation}
allowing the quadratic equation~(\ref{quadratic_isothermal}) to be written in the dimensionless form
\begin{equation}
  s^2=\hat\Sigma^2(1-s)
\end{equation}
in terms of the dimensionless parameter
\begin{equation}
  \hat\Sigma=\f{\mathcal{W}\pi G\Sigma}{c_\text{s}\nu}.
\end{equation}
Its solution,
\begin{equation}
  s=\f{\hat\Sigma}{2}\left(\sqrt{\hat\Sigma^2+4}-\hat\Sigma\right),
\end{equation}
is a monotonically increasing function of $\hat\Sigma$, with $\dd\ln s/\dd\ln\hat\Sigma=2(1-s)/(2-s)$. When self-gravity is weak, we have $s\approx\hat\Sigma\ll1$ and $H\approx c_\text{s}/\nu$. When self-gravity is strong, we have $1-s\approx\hat\Sigma^{-2}\ll1$ and $H\approx c_\text{s}^2/\mathcal{W}\pi G\Sigma$.

The explicit solution (\ref{h_s})--(\ref{pi_s}) simplifies, in the isothermal case, to
\begin{align}
  &H=\f{c_\text{s}}{\nu}\sqrt{1-s},\label{h_s_isothermal}\\
  &\Sigma=\f{c_\text{s}\nu}{\mathcal{W}\pi G}\f{s}{\sqrt{1-s}},\\
  &P=\f{c_\text{s}^3\nu}{\mathcal{W}\pi G}\f{s}{\sqrt{1-s}},\\
  &\Pi=\f{c_\text{s}^3\nu}{\mathcal{W}\pi G}\f{s(1+s)}{\sqrt{1-s}}.
\end{align}
Equation~(\ref{h_s_isothermal}) shows the extent to which the disc is compressed by self-gravity. The specific energy\footnote{Strictly speaking, the internal energy is replaced here by the variable part of the free energy of the isothermal system.} and enthalpy are
\begin{align}
  &E=c_\text{s}^2\left[\ln\left(\f{s}{1-s}\right)+\f{1}{2}(1+s)\right]+\cst,\\
  &\Upsilon=c_\text{s}^2\left[\ln\left(\f{s}{1-s}\right)+\f{3}{2}(1+s)\right]+\cst.
\end{align}

The effective sound speed (squared) is
\begin{equation}
  c^2=\f{\dd\Pi}{\dd\Sigma}=c_\text{s}^2\left(\f{2+3s-3s^2}{2-s}\right).
\end{equation}
As $s$ increases from $0$ to $1$, the quantity in brackets rises from $1$ to a maximum of $9-4\sqrt{3}\approx2.072$ at $s=2(1-1/\sqrt{3})\approx0.8453$, before decreasing to $2$ at $s=1$. The logarithmic derivative
\begin{equation}
  \Gamma^{(1)}=\f{\dd\ln\Pi}{\dd\ln\Sigma}=\f{2+3s-3s^2}{(2-s)(1+s)}
\end{equation}
rises from $1$ at $s=0$ to a peak of $11/9$ at $s=1/2$ before returning to $1$ at $s=1$. As for the higher logarithmic derivatives, as $s$ increases from $0$ to $1$, $\Gamma^{(2)}$ undulates from $0$ to $0.09$ to $-0.14$ to $0$, while $\Gamma^{(3)}$ undulates from $0$ to $0.04$ to $-0.19$ to $0.11$ to $0$.

The most important results here are that the effective sound speed of an isothermal disc can be significantly enhanced by self-gravity through the inclusion of the gravitational pressure, while the scaleheight can be significantly reduced by self-gravity.

\section{Affine dynamics of self-gravitating discs}
\label{s:affine}

In this section the results obtained so far in this paper for self-gravitating equilibria are extended and applied to dynamical situations. We do this by extending the affine model of the dynamics of astrophysical discs \citep{2018MNRAS.477.1744O} to include the energy and forces associated with self-gravity. We restrict our analysis to discs that are thin and are symmetric about the midplane, leaving to future work the study of warped self-gravitating discs.

\subsection{Lagrangian approach}

In the affine model, the disc is thought of as a continuous set of columnar fluid elements. We develop a Lagrangian description of the dynamics by considering a reference state that is most naturally assumed to be in equilibrium. Each column has a horizontal location in the reference state (for which we use a subscript zero) described by Cartesian coordinates $(x_0,y_0)$ in a global frame of reference, and a vertical structure given by a model such as the polytropic (or isothermal) discs described in Section~\ref{s:equilibrium}. The reference state has surface density $\Sigma_0$, scaleheight $H_0$ and vertically integrated pressure $P_0$, all depending in general on $(x_0,y_0)$.

We assume that the dynamical state of the disc is reached from the reference state by a time-dependent affine transformation of each vertical column, i.e.\ a combination of a translation and a linear transformation. In this paper we restrict the transformations to  preserve the reflectional symmetry about the midplane, which means that the columns are expanded or contracted but not tilted, and their centres remain in the plane $z=0$.

The dimensionless vertical coordinate $\zeta=z/H=z_0/H_0$ is used to label the fluid elements within each column, so it is a Lagrangian coordinate. The fluid element with coordinates $(x_0,y_0,z_0)=(x_0,y_0,H_0\zeta)$ in the reference state is found at $(x,y,z)=(X,Y,H\zeta)$ in the dynamical state at time $t$, where $X$, $Y$ and $H$ depend on $(x_0,y_0,t)$ but not on $\zeta$. This corresponds to an arbitrary horizontal relocation of the columns, combined with a vertical stretch by a factor of $H/H_0$.

In order to evaluate the density, pressure and internal energy of the disc we need the Jacobian determinant of the mapping from the reference state to the dynamical state. The Jacobian of the 3D transformation $(x_0,y_0,z_0)\mapsto(x,y,z)$ is
\begin{equation}
  J_3=\begin{vmatrix}\dfrac{\p X}{\p x_0}&\dfrac{\p X}{\p y_0}&0\\[8pt]\dfrac{\p Y}{\p x_0}&\dfrac{\p Y}{\p y_0}&0\\[8pt]z_0\dfrac{\p}{\p x_0}\left(\dfrac{H}{H_0}\right)&z_0\dfrac{\p}{\p y_0}\left(\dfrac{H}{H_0}\right)&\dfrac{H}{H_0}\end{vmatrix}=J_2\f{H}{H_0},
\end{equation}
where
\begin{equation}
  J_2=\begin{vmatrix}\dfrac{\p X}{\p x_0}&\dfrac{\p X}{\p y_0}\\[8pt]\dfrac{\p Y}{\p x_0}&\dfrac{\p Y}{\p y_0}\end{vmatrix}
\end{equation}
is the Jacobian determinant of the 2D transformation of the column centres in the midplane.

Given that mass and entropy are conserved under the transformation, the density and pressure (for a perfect gas of adiabatic exponent $\gamma$) in the dynamical state are then given by
\begin{equation}
  \rho=J_3^{-1}\rho_0,\qquad
  p=J_3^{-\gamma}p_0,
\end{equation}
while the surface density and vertically integrated pressure are
\begin{align}
  &\Sigma=J_3^{-1}\f{H}{H_0}\Sigma_0=J_2^{-1}\Sigma_0,\\
  &P=J_3^{-\gamma}\f{H}{H_0}P_0=J_2^{-\gamma}\left(\f{H}{H_0}\right)^{-(\gamma-1)}P_0.
\end{align}
The mass of the disc can be written variously as
\begin{align}
  \int\dd m&=\iint\Sigma_0\,\dd x_0\,\dd y_0=\iiint\rho_0\,\dd x_0\,\dd y_0\,\dd z_0\nonumber\\
  &=\iint\Sigma\,\dd x\,\dd y=\iiint\rho\,\dd x\,\dd y\,\dd z.
\end{align}

The Lagrangian of the ideal fluid is the kinetic energy minus the sum of gravitational and internal energies:
\begin{equation}
  \mathfrak{L}=\mathfrak{T}-\mathfrak{W}-\mathfrak{U}=\int\left(\f{1}{2}|\vecu|^2-\Phi_\text{ext}-\f{1}{2}\Phi_\text{int}-e\right)\dd m,
\end{equation}
where $\vecu$ is the velocity and $e=p/((\gamma-1)\rho)$. The conserved total energy of the fluid is $\mathfrak{T}+\mathfrak{W}+\mathfrak{U}$.

The velocity of a fluid element is given by
\begin{equation}
  \left(u_x,u_y,u_z\right)=\left(\f{\DD X}{\DD t},\f{\DD Y}{\DD t},\f{\DD H}{\DD t}\zeta\right),
\end{equation}
where we write $\DD/\DD t$ rather than $\p/\p t$ to emphasize that the time-derivative is taken in the Lagrangian sense, at fixed $(x_0,y_0)$. Note that $\zeta$ is a Lagrangian coordinate satisfying $\DD\zeta/\DD t=0$. Since the mass-weighted vertical average of $\zeta^2$ is unity, as follows from the definitions of $\Sigma$, $H$ and $\zeta$, the kinetic energy $\mathfrak{T}$ is
\begin{equation}
  \iint\f{1}{2}\left[\left(\f{\DD X}{\DD t}\right)^2+\left(\f{\DD Y}{\DD t}\right)^2+\left(\f{\DD H}{\DD t}\right)^2\right]\Sigma_0\,\dd x_0\,\dd y_0.
\end{equation}

The gravitational energy $\mathfrak{W}$ consists of several contributions. First, the energy of the (thin) disc in the external potential due to the central object is
\begin{equation}
  \mathfrak{W}_\text{ext}=\iint\left(\Phi_\text{c}+\f{1}{2}H^2\Psi_\text{c}\right)\Sigma_0\,\dd x_0\,\dd y_0,
\end{equation}
in which $\Phi_\text{c}$ and $\Psi_\text{c}$ are generally functions of $(X,Y,t)$.
Then the internal gravitational energy $\mathfrak{W}_\text{int}$ of the disc is expanded in the manner described in Section~\ref{s:expansion_gpe}, including far and near contributions. In this paper, we use a truncation of this expansion; we always include the dominant far and near contributions $\mathfrak{W}_1$ and $\mathfrak{W}_2$, and optionally include the quadrupolar term $\mathfrak{W}_3$ at our discretion.

Notably, the far contributions $\mathfrak{W}_1$ and $\mathfrak{W}_3$ are already expressed in terms of $\Sigma$ and $H$, without approximation (equations \ref{w1_integral} and \ref{w3_integral}). We have seen in Section~\ref{s:equilibrium} that the near contribution $\mathfrak{W}_2$ can be written in terms of $\Sigma$, $H$ and a dimensionless constant $\mathcal{W}$ in an almost universal way, independent of the detailed vertical structure of the disc. Thus we represent $\mathfrak{W}_2$ in the affine model as
\begin{equation}
  \int\f{W}{\Sigma}\,\dd m=\mathcal{W}\pi G\iint\Sigma H\,\Sigma_0\,\dd x_0\,\dd y_0,
\end{equation}
in which $\mathcal{W}$ is regarded as a constant.

Finally, the internal energy~$\mathfrak{U}$ is
\begin{equation}
  \int\f{P}{(\gamma-1)\Sigma}\,\dd m=\iint\f{J_3^{-(\gamma-1)}}{(\gamma-1)}P_0\,\dd x_0\,\dd y_0.
\end{equation}

The equations of motion for the ideal fluid can be derived by requiring the action $\int\mathfrak{L}\,\dd t$ to be stationary with respect to variations of the quantities $X$, $Y$ and $H$. We omit the details of this calculation, many of which are similar to those in \citet{2018MNRAS.477.1744O}.
Variation of the action with respect to $X$ and $Y$ leads to the horizontal equations of motion
\begin{align}
  &\f{\DD^2X}{\DD t^2}=-\f{\p\Phi}{\p X}-\f{1}{2}H^2\f{\p\Psi}{\p X}-\f{1}{\Sigma}\f{\p\Pi}{\p X},\\
  &\f{\DD^2Y}{\DD t^2}=-\f{\p\Phi}{\p Y}-\f{1}{2}H^2\f{\p\Psi}{\p Y}-\f{1}{\Sigma}\f{\p\Pi}{\p Y},
\end{align}
where
\begin{equation}
  \Pi=P+W=P+\mathcal{W}\pi G\Sigma^2H
\end{equation}
is the total pressure introduced in Section~\ref{s:equilibrium}, while variation of the action with respect to $H$ gives the dynamical equation for the scaleheight
\begin{equation}
  \f{\DD^2H}{\DD t^2}=-\Psi H-\mathcal{W}\pi G\Sigma+\f{P}{\Sigma H}.
\end{equation}
In these equations $\Phi=\Phi_\text{c}+\Phi_\text{d}$ combines the contributions from the central object and the far self-gravity of the disc, and similarly for $\Psi$, with (in the notation of Section~\ref{s:expansion_gpe})
\begin{align}
  &\Phi_\text{d}=-G\int\f{\Sigma'}{\Delta_2}\,\dd A'+G\int\f{\Sigma'H'^2}{2\Delta_2^3}\,\dd A',\label{phid}\\
  &\Psi_\text{d}=+G\int\f{\Sigma'}{\Delta_2^3}\,\dd A'.\label{psid}
\end{align}

\subsection{Eulerian form of the equations}
\label{s:eulerian}

In the Eulerian viewpoint we consider physical quantities as functions of the position $\barvecx=(x,y)$ in the plane, and time~$t$. By introducing the velocity components
\begin{equation}
  (v_x,v_y)=\left(\f{\DD X}{\DD t},\f{\DD Y}{\DD t}\right),\qquad
  w=\f{\DD H}{\DD t}
\end{equation}
and the horizontal gradient operator
\begin{equation}
  \gradbar=\left(\f{\p}{\p x},\f{\p}{\p y}\right)
\end{equation}
as in section~10 of \citet{2018MNRAS.477.1744O}, we can also write our equations of motion in the Eulerian form
\begin{align}
  &\left(\f{\p}{\p t}+\vecv\bcdot\gradbar\right)\vecv=-\gradbar\Phi-\f{1}{2}H^2\gradbar\Psi-\f{1}{\Sigma}\gradbar\Pi,\label{eulerian_v}\\
  &\left(\f{\p}{\p t}+\vecv\bcdot\gradbar\right)w=-H\Psi-\mathcal{W}\pi G\Sigma+\f{P}{\Sigma H},\label{eulerian_w}\\
  &\left(\f{\p}{\p t}+\vecv\bcdot\gradbar\right)H=w,\label{eulerian_h}
\end{align}
together with an equation of mass conservation,
\begin{equation}
  \left(\f{\p}{\p t}+\vecv\bcdot\gradbar\right)\Sigma=-\Sigma\gradbar\bcdot\vecv,\label{eulerian_sigma}
\end{equation}
and a thermal energy equation, for example in the form
\begin{equation}
  \left(\f{\p}{\p t}+\vecv\bcdot\gradbar\right)P=-\gamma P\gradbar\bcdot\vecv-\f{(\gamma-1)Pw}{H}\label{eulerian_p}
\end{equation}
(see section~7 of \citealt{2018MNRAS.477.1744O} for alternatives).

\subsection{Conservation laws}

In the absence of far self-gravity, the energy conservation equation is
\begin{equation}
  \f{\p}{\p t}\left(\Sigma\mathcal{E}\right)+\grad\bcdot\left[\left(\Sigma\mathcal{E}+\Pi\right)\vecv\right]=\Sigma\left(\f{\p\Phi_\text{c}}{\p t}+\f{1}{2}H^2\f{\p\Psi_\text{c}}{\p t}\right),
\end{equation}
where
\begin{equation}
  \mathcal{E}=\f{1}{2}\left(|\vecv|^2+w^2\right)+\Phi_\text{c}+E,
\end{equation}
with
\begin{equation}
  E=\f{1}{2}H^2\Psi_\text{c}+\f{P}{(\gamma-1)\Sigma}+\mathcal{W}\pi G\Sigma H,
\end{equation}
consistent with Section~\ref{s:equilibrium}. Energy is conserved unless the external potential depends on time.

The additional energy associated with far self-gravity is the sum of $\mathfrak{W}_1$ and $\mathfrak{W}_3$ (equations \ref{w1_integral} and \ref{w3_integral}), i.e.
\begin{equation}
  -G\iint\f{\Sigma\Sigma'}{2\Delta_2}\,\dd A\,\dd A'+G\iint\f{\Sigma\Sigma'\left(H^2+H'^2\right)}{4\Delta_2^3}\,\dd A\,\dd A',
\end{equation}
and its time derivative is
\begin{align}
  &G\iint\f{\gradbar\bcdot\left(\Sigma\vecv\right)\Sigma'}{\Delta_2}\,\dd A\,\dd A'\nonumber\\
  &\qquad-G\iint\f{\gradbar\bcdot\left(\Sigma\vecv\right)\Sigma'\left(H^2+H'^2\right)}{2\Delta_2^3}\,\dd A\,\dd A'\nonumber\\
  &\qquad+G\iint\f{\Sigma\Sigma'H\left(w-\vecv\bcdot\gradbar H\right)}{\Delta_2^3}\,\dd A\,\dd A',
\end{align}
which can be integrated to give a boundary term plus
\begin{equation}
  \int\left[\Sigma\vecv\bcdot\left(\gradbar\Phi_\text{d}+\f{1}{2}H^2\gradbar\Psi_\text{d}\right)+\Sigma wH\Psi_\text{d}\right]\dd A.
\end{equation}
These are precisely the missing terms required to construct a total energy equation in conservative form.

Equations (\ref{eulerian_v})--(\ref{eulerian_p}) also imply the conservation of entropy and potential vorticity (PV) in the form
\begin{equation}
  \f{\DD}{\DD t}\ln\left(\f{PH^{\gamma-1}}{\Sigma^\gamma}\right)=0,\qquad
  \f{\DD q}{\DD t}=\f{S_q}{\Sigma},
\end{equation}
with PV
\begin{equation}
  q=\f{1}{\Sigma}\left[\f{\p}{\p x}\left(v_y+w\f{\p H}{\p y}\right)-\f{\p}{\p y}\left(v_x+w\f{\p H}{\p x}\right)\right]
\end{equation}
and baroclinic source term
\begin{equation}
  S_q=\f{\p}{\p x}\left(\f{P}{H}\right)\f{\p}{\p y}\left(\f{H}{\Sigma}\right)-\f{\p}{\p y}\left(\f{P}{H}\right)\f{\p}{\p x}\left(\f{H}{\Sigma}\right),
\end{equation}
as in section~8 of \citet{2018MNRAS.477.1744O}, unaffected by self-gravity. This is to be expected because the gravitational force is irrotational.

\subsection{Local approximation}
\label{s:local}

In the local approximation, also known as the shearing sheet or shearing box \citep{1965MNRAS.130..125G,1995ApJ...440..742H,2017MNRAS.472.1432L}, we represent a small patch of the disc in a local Cartesian coordinate system centred on a reference point in a circular orbit in the midplane, and rotating with that orbit.

The constant angular velocity $\Omega$ of the local model is that of the reference orbit, taking into account the combined potentials of the central object and the disc as a whole (assumed axisymmetric).

In order to be able to treat self-gravitating structures such as density waves within the local model, we need to allow part of the far self-gravity of the disc to have a local contribution, which results from variations in the surface density of the matter contained within the scope of the model about the reference value of the surface density.

The Eulerian form of the equations in the local model is the same as given in Section~\ref{s:eulerian}, except for the inclusion of the Coriolis force in the horizontal equation of motion:
\begin{equation}
  \left(\f{\p}{\p t}+\vecv\bcdot\gradbar\right)\vecv+2\Omega\,\vece_z\times\vecv=-\gradbar\Phi-\f{1}{2}H^2\gradbar\Psi-\f{1}{\Sigma}\gradbar\Pi,
\end{equation}
and a different representation of the gravitational potentials $\Phi$ and~$\Psi$. When the potential in the midplane due to the central object and the disc as a whole is combined with the centrifugal potential due to the rotation of the frame of reference and expanded to second order in the local radial coordinate $x$, it results in a tidal potential $\Phi_\text{t}=-\Omega Sx^2$. The gradient of $\Phi_\text{t}$ balances the Coriolis force of the orbital shear flow $v_y=-Sx$, which is the local representation of the family of circular orbits in the midplane. Then the total potential in the midplane in the local model is $\Phi=\Phi_\text{t}+\Phi_\text{d}$, where 
\begin{equation}
  \Phi_\text{d}=-G\int\f{\Sigma'-\Sigma_0}{\Delta_2}\,\dd A'+G\int\f{\Sigma'H'^2-\Sigma_0H_0^2}{2\Delta_2^3}\,\dd A'\label{phid_local}
\end{equation}
and $\Sigma_0$ and $H_0$ are the reference surface density and scaleheight of the local model. Similarly $\Psi=\nu^2+\Psi_\text{d}$, where $\nu$ is the vertical oscillation frequency due to the central object and the disc as a whole, and
\begin{equation}
  \Psi_\text{d}=+G\int\f{\Sigma'-\Sigma_0}{\Delta_2^3}\,\dd A'.\label{psid_local}
\end{equation}

\subsection{Interpretation of the gravitational pressure}

We have seen that the near self-gravity of the disc contributes both to its vertical equation of motion (e.g.\ equation~\ref{eulerian_w}), through the term $-W/\Sigma H$ that acts to compress the disc, and to its horizontal equation of motion (e.g.\ equation~\ref{eulerian_v}), through the term $-(\bar\grad W)/\Sigma$ that acts similarly to a pressure in the plane of the disc. Both effects were shown above to derive from the potential energy contribution $W\propto\Sigma^2H$ per unit area. The gravitational pressure results from the dependence of $W$ on $\Sigma$, and behaves similarly to a (2D) gas with adiabatic exponent~$2$.

The gravitational pressure can also be explained using the gravitational stress tensor \citep{1970RSPSA.320..277M,1972MNRAS.157....1L}, which is commonly used to quantify the transport of angular momentum  by gravitational turbulence in discs \citep[e.g.][]{2001ApJ...553..174G}. The self-gravitational force per unit volume, $\rho\vecg$, can be written as the divergence of the stress tensor
\begin{equation}
  G_{ij}=-\f{g_ig_j}{4\pi G}+\f{|\vecg|^2}{8\pi G}\,\delta_{ij},
\end{equation}
using Poisson's equation $\grad\bcdot\vecg=-4\pi G\rho$ and the property $\grad\times\vecg=\veczero$.

In a horizontally uniform equilibrium as considered in Section~\ref{s:equilibrium}, $\vecg$ has only a $z$-component that decreases from $+2\pi G\Sigma$ below the disc to $-2\pi G\Sigma$ above it, vanishing at the midplane. Then
\begin{equation}
  G_{zz}=-\f{g^2}{8\pi G}
\end{equation}
has a maximum in the midplane, while
\begin{equation}
  G_{xx}=G_{yy}=+\f{g^2}{8\pi G}
\end{equation}
has a minimum in the midplane. Therefore $G_{zz}$ acts oppositely to a pressure in the vertical direction, tending to compress the disc, while $G_{xx}$ and $G_{yy}$ act similarly to a pressure in the plane of the disc (if we recall that a pressure corresponds to a negative isotropic stress). After vertical integration, the `deficit' in $G_{xx}$ and $G_{yy}$ due to the depletion of $g^2$ near the midplane can be identified precisely with the gravitational pressure $W$.

Although in such a horizontally uniform equilibrium the gravitational pressure is also uniform and has no effect on the dynamics, it will play a role when horizontal variations are introduced, as we consider next.

\section{Axisymmetric waves and linear stability}
\label{s:stability}

In this section we derive the dispersion relation for axisymmetric perturbations of the disc, using the affine model in the local approximation. This is appropriate for wavelengths that are longer than the scaleheight $H$ but small compared to the radius $r$. We obtain the conditions for marginal stability of the disc and show how the Toomre stability parameter can be usefully redefined. We then compare the results of the affine model with numerical results on marginal stability for a range of isentropic, isothermal and incompressible disc models.

\subsection{Linear perturbations}

Using the local affine model of Section~\ref{s:local}, we consider a
horizontally uniform basic state with surface density $\Sigma=\Sigma_0$, scaleheight $H=H_0$ and vertically integrated pressure $P=P_0$, and with the orbital shear flow $v_y=-Sx$, but otherwise static ($v_x=w=0$). Vertical equilibrium requires
\begin{equation}
  P=\nu^2\Sigma H^2+\mathcal{W}\pi G\Sigma^2H,
\label{hydrostatic_integrated}
\end{equation}
as considered in Section~\ref{s:equilibrium}.

We then introduce axisymmetric linear perturbations $\propto\exp\left(\ii kx-\ii\omega t\right)$, denoted by the symbol~$\delta$. The linearized equations of Section~\ref{s:local} are
\begin{align}
  &-\ii\omega\,\delta v_x-2\Omega\,\delta v_y=-\ii k\left(\delta\Phi_\text{d}+\f{1}{2}H^2\,\delta\Psi_\text{d}+\f{\delta\Pi}{\Sigma}\right),\label{linearized1}\\
  &-\ii\omega\,\delta v_y+(2\Omega-S)\,\delta v_x=0,\\
  &-\ii\omega\,\delta w=-\nu^2\,\delta H-H\,\delta\Psi_\text{d}-\mathcal{W}\pi G\,\delta\Sigma\nonumber\\
  &\qquad\qquad\qquad+\f{P}{\Sigma H}\left(\f{\delta P}{P}-\f{\delta\Sigma}{\Sigma}-\f{\delta H}{H}\right),\\
  &-\ii\omega\,\delta H=\delta w,\\
  &-\ii\omega\,\delta\Sigma=-\Sigma\,\ii k\,\delta v_x,\\
  &-\ii\omega\,\delta P=-\gamma P\,\ii k\,\delta v_x-\f{(\gamma-1)P\,\delta w}{H},\label{linearized6}
\end{align}
together with
\begin{align}
  &\delta\Phi_\text{d}=-\f{2\pi G\Sigma}{|k|}\left[\f{\delta\Sigma}{\Sigma}+(kH)^2\left(\f{1}{2}\f{\delta\Sigma}{\Sigma}+\f{\delta H}{H}\right)\right],\\
  &\delta\Psi_\text{d}=-2\pi G\Sigma|k|\f{\delta\Sigma}{\Sigma},\\
  &\delta\Pi=\delta P+W\left(\f{2\,\delta\Sigma}{\Sigma}+\f{\delta H}{H}\right).
\end{align}

To see where the expressions for $\delta\Phi_\text{d}$ and $\delta\Psi_\text{d}$ come from, we note that the 2D Fourier transforms of $1/r$ and $1/r^3$ are $2\pi/|k|$ and $-2\pi|k|$, respectively, where $r=\sqrt{x^2+y^2}$ and $|k|=\sqrt{k_x^2+k_y^2}$. Interpreting the integrals in equations (\ref{phid_local}) and (\ref{psid_local}) as convolutions leads to the expressions for $\delta\Phi_\text{d}$ and $\delta\Psi_\text{d}$ above.

We can combine equations (\ref{linearized1})--(\ref{linearized6}) to obtain the linearized forms of the conservation laws for entropy and PV:
\begin{align}
  &-\ii\omega\left[\f{\delta P}{P}+(\gamma-1)\f{\delta H}{H}-\gamma\f{\delta\Sigma}{\Sigma}\right]=0,\label{conservation1}\\
  &-\ii\omega\left[\f{\ii k\,\delta v_y}{\Sigma}-(2\Omega-S)\f{\delta\Sigma}{\Sigma^2}\right]=0.\label{conservation2}
\end{align}
There exist zero-frequency modes ($\omega=0$) for which the quantities in square brackets can be non-zero. These solutions represent small-amplitude zonal flows in which there are stationary, sinusoidal variations of $v_y$, $\Sigma$, $P$, $H$, $\Phi_\text{d}$ and $\Psi_\text{d}$. The linearized equations in the case $\omega=0$, $\delta v_x=0$ impose three algebraic constraints on these variations, leaving two degrees of freedom, because the zonal flows can have arbitrary small sinusoidal variations of entropy and PV \citep[cf.][]{2016MNRAS.463.3725V}.

In the remaining modes (including all non-zero frequency modes), the quantities in square brackets in equations (\ref{conservation1}) and (\ref{conservation2}) are zero, meaning that the perturbations are isentropic and isovortical. The linearized equations can be combined to deduce a somewhat complicated dispersion relation that describes both density waves and a higher-frequency branch of acoustic waves.

To investigate the conditions for marginal GI, we consider isentropic and isovortical perturbations and then set $\omega^2$ to zero, leading to
\begin{align}
  &\kappa^2\f{\delta\Sigma}{\Sigma}+k^2\left(\delta\Phi_\text{d}+\f{1}{2}H^2\,\delta\Psi_\text{d}+\f{\delta\Pi}{\Sigma}\right)=0,\label{kasqdsos}\\
  &\nu^2\,\delta H+H\,\delta\Psi_\text{d}+\mathcal{W}\pi G\,\delta\Sigma=\f{P}{\Sigma H}\left(\f{\delta P}{P}-\f{\delta\Sigma}{\Sigma}-\f{\delta H}{H}\right),\label{nusqdh}\\
  &\f{\delta P}{P}+(\gamma-1)\f{\delta H}{H}-\gamma\f{\delta\Sigma}{\Sigma}=0,\label{dpop}\\
  &\delta\Pi=\delta P+W\left(\f{2\,\delta\Sigma}{\Sigma}+\f{\delta H}{H}\right).
\end{align}
The meaning of these equations is that we are seeking a nearby equilibrium state of the disc, which has the same uniform entropy and PV as the basic state but has small-amplitude sinusoidal variations of $v_y$, $\Sigma$, $P$, $H$, $\Phi_\text{d}$ and $\Psi_\text{d}$.

If it were not for the $\delta\Psi_\text{d}$ term, equation~(\ref{nusqdh}) would represent a purely local linearization of the hydrostatic condition~(\ref{hydrostatic_integrated}) for each column, which when combined with equation~(\ref{dpop}) would correspond to a small displacement along the isentropic equilibrium sequence considered in Section~\ref{s:isentropic}. Then $\delta\Pi$ and $\delta\Sigma$ would be related via $\delta\Pi=c^2\,\delta\Sigma$, where $c^2=\dd\Pi/\dd\Sigma$ along this sequence. The $\delta\Psi_\text{d}$ term takes the perturbation slightly away from the equilibrium sequence, such that
\begin{align}
  \delta\Pi-c^2\,\delta\Sigma&=\left[\f{(\gamma-1)P-W}{\f{\gamma P}{\Sigma H^2}+\nu^2}\right]\,\delta\Psi_\text{d}\nonumber\\
  &=\left(\f{\gamma-1-s}{\gamma+1-s}\right)\Sigma H^2\,\delta\Psi_\text{d}.
\end{align}
When all terms in equation~(\ref{kasqdsos}) are reexpressed in terms of $\delta\Sigma/\Sigma$, we obtain the condition for marginal stability in the quartic form
\begin{align}
  &\kappa^2-2\pi G\Sigma|k|\left[1+\left(\f{3\gamma-1-3s}{\gamma+1-s}\right)(kH)^2\right]+c^2k^2\nonumber\\
  &\qquad-\f{(2\pi G\Sigma)^2(kH)^4}{\f{\gamma P}{\Sigma}+\nu^2H^2}=0.
\label{marginal_quartic}
\end{align}
The quadrupolar effects are the correction factor in square brackets and the last, quartic, term.

Let us rewrite the marginal condition in the dimensionless form
\begin{align}
  \f{\kappa^2}{\nu^2}=&\f{(2/\mathcal{W})s}{(1-s)}|kH|\left[1+\f{(3\gamma-1-3s)|kH|^2+(2/\mathcal{W})s|kH|^3}{\gamma+1-s}\right]\nonumber\\
  &-\f{(3\gamma-1+3\gamma s-3s^2)}{(1-s)(\gamma+1-s)}|kH|^2.
\label{marginal_quartic_dimensionless}
\end{align}
We can then truncate the quartic on the RHS at different orders and set the discriminant to zero to locate the marginal equilibrium for various values of $\kappa^2/\nu^2$ and $\gamma$. The results can be compared with exact analytical results for marginal stability in the cases $n=0$ and $n=1$ (Appendix~\ref{s:appendix_marginal}) and with numerical results for other values of~$n$ \citep{2025arXiv250111658B}.

\begin{figure*}
  \includegraphics[width=0.68\columnwidth]{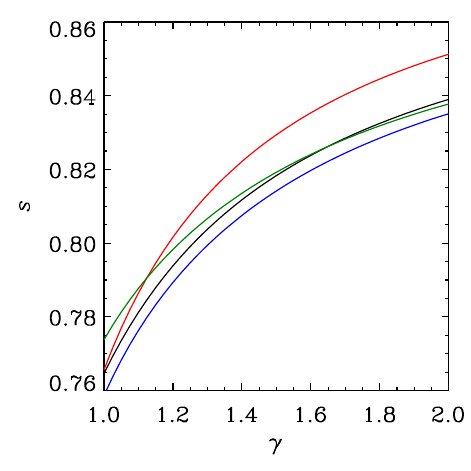}
  \includegraphics[width=0.68\columnwidth]{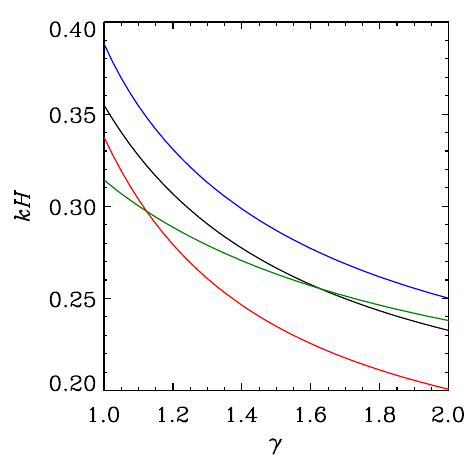}
  \includegraphics[width=0.68\columnwidth]{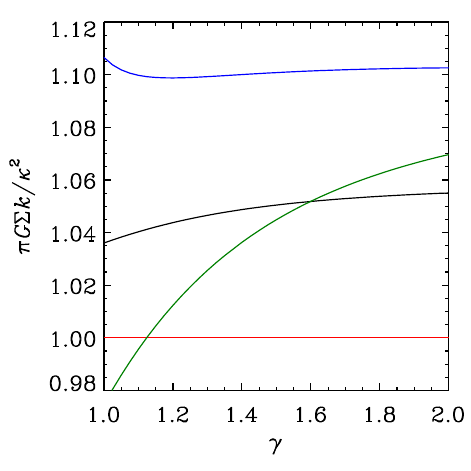}
  \caption{Dimensionless characterization of isentropic discs that are marginally stable to the axisymmetric GI. The degree of self-gravity $s$ (left), the radial wavenumber in `scaleheight units' (middle) and the radial wavenumber in `Toomre units' (right) are plotted against the adiabatic exponent $\gamma$ in the case $\kappa^2/\nu^2=1$. The black lines are numerical results for 3D discs \citep{2025arXiv250111658B}, while the coloured lines are the results of the affine model in quadratic (red), cubic (green) and quartic (blue) approximations.}
\label{f:marginal}
\end{figure*}

This comparison (Fig.~\ref{f:marginal}) shows that quadratic truncation (in which quadrupolar self-gravity is neglected) gives a good approximation to marginal stability. Cubic truncation (which includes quadrupolar self-gravity) gives a better approximation, except for $\gamma$ close to~$1$. The full quartic, however, generally gives a poorer approximation than the cubic truncation. The reason for this appears to be that the quartic term in equation~(\ref{marginal_quartic}) results from the combination of two quadrupolar effects,\footnote{A change in surface density at one location affects (through $\delta\Psi_\text{d}$) the scaleheight ($\delta H$) at a second location. This in turn affects the potential ($\delta\Phi_\text{d}$) at a third location.} but the quartic coefficient is very inaccurate (even having the wrong sign) because it is incomplete, missing higher-order
self-gravitational effects that are at least as important.

\subsection{Neglecting quadrupolar self-gravity}

If the effect of quadrupolar self-gravity on the perturbations is neglected by omitting $\delta\Psi_\text{d}$ and simplifying $\delta\Phi_\text{d}$ to the usual expression
\begin{equation}
  \delta\Phi_\text{d}=-\f{2\pi G\Sigma}{|k|}\f{\delta\Sigma}{\Sigma},
\end{equation}
then the dispersion relation (described, but not written in full in the previous subsection) simplifies to
\begin{align}
  &\left[\omega^2-\kappa^2+2\pi G\Sigma|k|-\left(\f{\gamma P}{\Sigma}+2\mathcal{W}\pi G\Sigma H\right)k^2\right]\nonumber\\
  &\quad\times\left(\omega^2-\nu^2-\f{\gamma P}{\Sigma H^2}\right)=\left[(\gamma-1)\f{P}{\Sigma H}-\mathcal{W}\pi G\Sigma\right]^2k^2,
\end{align}
which generalizes equation~142 of \citet{2018MNRAS.477.1744O} to include self-gravity. Note that the gravitational pressure $W=\mathcal{W}\pi G\Sigma^2H$ appears on the left-hand side in a way similar to the pressure of a perfect gas with $\gamma=2$, which results from its proportionality to $\Sigma^2$.

Using $c^2=\dd\Pi/\dd\Sigma$, the dispersion relation can also be written as
\begin{align}
  &\omega^4-\omega^2\left[\kappa^2-2\pi G\Sigma|k|+\left(\f{\gamma P+2W}{\Sigma}\right)k^2+\nu^2+\f{\gamma P}{\Sigma H^2}\right]\nonumber\\
  &\qquad+\left(\nu^2+\f{\gamma P}{\Sigma H^2}\right)\left(\kappa^2-2\pi G\Sigma|k|+c^2k^2\right)=0.
\end{align}
This is a quadratic equation for $\omega^2$ and describes two wave modes. The lower branch is the density wave, which becomes unstable ($\omega^2<0$) for intermediate wavenumbers if the self-gravity is sufficiently strong. The upper branch is a breathing or acoustic mode, which is stable. \citet{2018MNRAS.477.1744O} found that the dispersion relation for a non-self-gravitating disc is inaccurate for $kH\gtrsim1$ unless the affine model is modified.

Setting $\omega^2$ to zero, we obtain the condition for marginal stability (corresponding to the red lines in Fig.~\ref{f:marginal}). This is a quadratic equation for $|k|$,
\begin{equation}
  \kappa^2-2\pi G\Sigma|k|+c^2k^2=0,
\label{marginal_monopolar}
\end{equation}
and is formally identical to that obtained for a 2D (razor-thin) disc. Stability is determined by how the Toomre parameter
\begin{equation}
  Q=\f{\kappa c}{\pi G\Sigma}
\label{q_new}
\end{equation}
compares with the critical value of~$1$. If $Q>1$ then the marginal stability condition~(\ref{marginal_monopolar}) cannot be satisfied for any $|k|$ and the disc is stable ($\omega^2>0$ for all real wavenumbers). If $Q<1$ then the quadratic equation for $|k|$ has two positive roots bracketing a band of wavenumbers on which the disc is unstable. Marginal stability occurs at $Q=1$ with the critical wavenumber
\begin{equation}
  k=\f{\kappa^2}{\pi G\Sigma}=\f{\kappa}{c},
\label{k_crit}
\end{equation}
as in the 2D theory of GI.

The essential difference between the affine theory of GI, in the monopolar approximation adopted here, and the 2D theory is that the effective sound speed $c$ is enhanced by the gravitational pressure in addition to the gas pressure. (For example, we have seen in the case of isothermal discs in Section~\ref{s:isothermal} that this effect can approximately double the value of $c^2$.) Furthermore, the compression of the disc by near self-gravity means that the critical wavenumber~(\ref{k_crit}) can be much less than the value $1/H$ given by the 2D theory of an isothermal disc, meaning that the critical wavelength can be much longer than $2\pi H$.

Using equation~(\ref{csq_isentropic}) and the equilibrium relations, the marginal disc condition $Q=1$ can be written as
\begin{equation}
  \f{3\gamma-1+3\gamma s-3s^2}{\gamma+1-s}=\f{s^2}{1-s}\f{1}{\mathcal{W}^2}\f{\nu^2}{\kappa^2}.
\end{equation}
For given values of $\gamma$ and $\nu^2/\kappa^2$, this equation can be solved to locate the critical equilibrium $s=s_\text{crit}$ along an isentropic sequence. (In principle, $\mathcal{W}$ varies slightly with $s$, but we have seen that it is a very good approximation to regard $\mathcal{W}$ as independent of $s$.)

In the isothermal case $\gamma=1$, the usual Toomre parameter would be defined by
\begin{equation}
  Q_\text{iso}=\f{\kappa c_\text{s}}{\pi G\Sigma}.
\end{equation}
According to the affine model, neglecting quadrupolar gravity, the marginally stable disc has $Q=1$, where $Q$ is defined by equation~(\ref{q_new}). This differs from $Q_\text{iso}$ by a factor of $c/c_\text{s}$, which (as discussed in Section~\ref{s:isothermal}) is approximately $1.43$ for a marginal equilibrium, leading to a critical value of $Q_\text{iso}$ of approximately $0.70$. The accurate critical value of $Q_\text{iso}$ derived from our numerical calculations \citep{2025arXiv250111658B} is $0.7062$.

In their review article, \citet{2016ARA&A..54..271K} present an interesting argument that leads to a qualitatively similar conclusion. They state that the non-zero thickness of the disc dilutes the term $-2\pi G\Sigma|k|$ in the dispersion relation by a factor of $\exp(-|k|H)$. When expanded to first order in $|k|H$, this results in a `pressure-like' correction $+2\pi G\Sigma Hk^2$ that combines with the term $c_\text{s}^2k^2$ and, when combined with an approximate treatment of  hydrostatic equilibrium, leads to a critical $Q_\text{iso}$ of about $0.6$. This argument can be seen as a hint of the gravitational pressure, which we have placed on a firmer theoretical basis. (Our equivalent expression for the correction term would also include a factor of $\mathcal{W}$.) The idea that the non-zero thickness of the disc weakens its self-gravity and thereby increases the surface density needed for instability was already stated by \citet{1964ApJ...139.1217T}.

\section{Weakly nonlinear gravitational instability}
\label{s:nonlinear}

\citet{2022ApJ...934L..19D} investigated the nonlinear dynamics of GI in an ideal fluid disc, using a 2D disc model in which the vertically integrated pressure and density are related by $P\propto\Sigma^\Gamma$ for some $\Gamma\ge1$. They found that the axisymmetric GI is subcritical for $\Gamma<5/3$ (or for $\Gamma>2$), meaning that the weakly nonlinear effects are destabilizing and allow interesting dynamics to occur in the linearly stable regime $Q>1$. \citet{2022ApJ...934L..19D} computed radially periodic nonlinear axisymmetric equilibria that emerge from a bifurcation at $Q=1$ and exist on the linearly stable side $Q>1$. They found that these equilibria can take the form of solitary waves, independent of boundary conditions, in a large domain and that they can be linearly unstable to non-axisymmetric disturbances in the form of trailing spiral waves.

We recently discovered the remarkable work of \citet{1979SvA....23..153M} \citep[see also][]{1984pgs2.book.....F} in which a related nonlinear dynamics was discussed, and the same type of solitary waves were described in a long-wavelength limit.

Following \citet{2022ApJ...934L..19D}, we consider nonlinear axisymmetric equilibria in the local approximation, but using the affine model (omitting quadrupolar self-gravity) to represent 3D effects. These equilibria satisfy radial and vertical force balance:
\begin{align}
  &-2\Omega v_y=2\Omega Sx-\f{\p\Phi_\text{d}}{\p x}-\f{1}{\Sigma}\f{\p\Pi}{\p x},\label{nonlinear1}\\
  &0=-\nu^2H-\mathcal{W}\pi G\Sigma+\f{P}{\Sigma H},\label{nonlinear2}
\end{align}
with $\Pi=P+\mathcal{W}\pi G\Sigma^2H$, and have the same uniform entropy and PV as in the uniform reference state:
\begin{align}
  &\f{P}{H}\left(\f{\Sigma}{H}\right)^{-\gamma}=\f{P_0}{H_0}\left(\f{\Sigma_0}{H_0}\right)^{-\gamma}=K_2,\label{nonlinear3}\\
  &\f{2\Omega+\f{\p v_y}{\p x}}{\Sigma}=\f{2\Omega-S}{\Sigma_0},\label{nonlinear4}
\end{align}
as well as the same mean surface density.

As in the study of uniform equilibria, the algebraic equations (\ref{nonlinear2}) and (\ref{nonlinear3}) defining a hydrostatic disc with a given entropy can be solved to determine $\Pi$ as a function of $\Sigma$.
Using the specific enthalpy $\Upsilon(\Sigma)$ such that $\dd\Pi=\Sigma\,\dd\Upsilon$, and eliminating $v_y$ between the remaining equations (\ref{nonlinear1}) and (\ref{nonlinear4}), we obtain
\begin{equation}
  \kappa^2\sigma=\p_x^2(\Phi_\text{d}+\Upsilon),\label{nonlinear5}
\end{equation}
where
\begin{equation}
  \sigma=\f{\Sigma}{\Sigma_0}-1
\end{equation}
is the fractional surface density perturbation
and, in the Fourier domain,
\begin{equation}
  \tilde\Phi_\text{d}=-\f{2\pi G\Sigma_0}{|k|}\tilde\sigma.
\end{equation}
Equation~(\ref{nonlinear5}) is formally equivalent to equation~5 of \citet{2022ApJ...934L..19D}, but the relation between enthalpy and density is different from the pure power law assumed there (see Section~\ref{s:isentropic}).

To develop a weakly nonlinear theory of axisymmetric GI, we expand the specific enthalpy in powers of the surface density perturbation,
\begin{equation}
  \Upsilon=\Upsilon_0+\Upsilon_1\sigma+\Upsilon_2\sigma^2+\Upsilon_3\sigma^3+\cdots,
\end{equation}
where
\begin{equation}
  \Upsilon_j=\f{\Sigma_0^j}{j!}\left(\f{\dd^j\Upsilon}{\dd\Sigma^j}\right)_0.
\end{equation}

For equation~(\ref{nonlinear5}) to have a solution $\sigma\propto\cos(kx)$ of infinitesimal amplitude and wavenumber $k$ would require the condition of marginal linear stability,
\begin{equation}
  \kappa^2=2\pi G\Sigma_0|k|-c^2k^2,
\label{marginal}
\end{equation}
with $c^2=\Upsilon_1$, to be satisfied. Optimization with respect to $|k|$ leads to
\begin{equation}
  |k|=\f{\pi G\Sigma_0}{c^2}=\f{\kappa^2}{\pi G\Sigma_0}
\label{marginal_optimized}
\end{equation}
and $Q=\kappa c/\pi G\Sigma_0$=1. For given values of $K_2$, $\nu$ and $\kappa$, the marginal condition $Q=1$ is satisfied for some surface density $\Sigma_\text{m}$ and wavenumber $|k|=k_\text{m}$.

In a weakly nonlinear theory, we seek periodic solutions
\begin{equation}
  \sigma=\sum_{n=1}^\infty\sigma_n\cos(nkx),
\end{equation}
in which the Fourier coefficients have the expansions
\begin{equation}
\begin{split}
  \sigma_1&=\epsilon\sigma_{11}+\epsilon^3\sigma_{13}+\cdots,\\
  \sigma_2&=\epsilon^2\sigma_{22}+\cdots,\\
  \sigma_3&=\epsilon^3\sigma_{33}+\cdots,
\end{split}
\end{equation}
etc., where $\epsilon\ll1$ is a small parameter measuring the amplitude of the primary component $\propto\cos(kx)$, and the higher-order terms are generated by the nonlinear relation between $\Upsilon$ and $\sigma$.

We allow for the surface density and wavenumber to differ slightly from their marginal values according to linear theory:
\begin{align}
  &\Sigma_0=\Sigma_\text{m}+\epsilon^2\Sigma_2+\cdots,\\
  &k=k_\text{m}+\epsilon k_1+\cdots.
\end{align}
This also means that
\begin{equation}
  c=c_\text{m}+\epsilon^2c_2+\cdots,
\end{equation}
where $c_2=(\dd c/\dd\Sigma)_\text{m}\Sigma_2$.
Substituting these expansions into equation~(\ref{nonlinear5}) and equating coefficients of successive powers of $\epsilon$, we find
\begin{align}
  &\kappa^2\sigma_{11}=2\pi G\Sigma_\text{m}k_\text{m}\sigma_{11}-k_\text{m}^2c_\text{m}^2\sigma_{11},\label{wnl1}\\
  &\kappa^2\sigma_{22}=2\pi G\Sigma_\text{m}(2k_\text{m})\sigma_{22}-(2k_\text{m})^2\left(c_\text{m}^2\sigma_{22}+\f{1}{2}\Upsilon_2\sigma_{11}^2\right),\label{wnl2}\\
  &\kappa^2\sigma_{13}=2\pi G\Sigma_\text{m}k_\text{m}\sigma_{13}+2\pi G\Sigma_2k_\text{m}\sigma_{11}\nonumber\\
  &\qquad\qquad-k_\text{m}^2\left(c_\text{m}^2\sigma_{13}+\Upsilon_2\sigma_{11}\sigma_{22}+\f{3}{4}\Upsilon_3\sigma_{11}^3\right)\nonumber\\
  &\qquad\qquad-k_1^2c_\text{m}^2\sigma_{11}-k_\text{m}^2(2c_\text{m}c_2)\sigma_{11},\label{wnl3}
\end{align}
in the last of which we have projected equation~(\ref{nonlinear5}) onto the $n=1$ mode only. (Terms linear in $k_1$ cancel because of equation~\ref{marginal_optimized}.)

Equation~(\ref{wnl1}) is satisfied for any $\sigma_{11}$, because of the marginal stability condition~(\ref{marginal}), and the corresponding terms involving $\sigma_{13}$ in equation~(\ref{wnl3}) cancel for the same reason. Solving equation~(\ref{wnl2}) for $\sigma_{22}$ and substituting into equation~(\ref{wnl3}), we obtain
\begin{align}
  &\sigma_{22}=-\f{2\Upsilon_2}{\Upsilon_1}\sigma_{11}^2,\\
  &\left(\f{3\Upsilon_3}{8\Upsilon_1}-\f{\Upsilon_2^2}{\Upsilon_1^2}\right)\sigma_{11}^2=\left(\f{\Sigma_2}{\Sigma_\text{m}}-\f{c_2}{c_\text{m}}\right)-\f{1}{2}\f{k_1^2}{k_\text{m}^2}.\label{wnl_amplitude}
\end{align}
The combination in brackets on the right-hand side of equation~(\ref{wnl_amplitude}) can be written as $(Q^{-1})_2$, since $Q^{-1}\propto\Sigma/c$ and its value at marginal stability is $Q_\text{m}=1$. It is a measure of the linear stability of the system. If the expansion coefficients are evaluated at marginal stability, then
\begin{equation}
  (Q^{-1})_2=\f{\Sigma_2}{\Sigma_\text{m}}-\f{c_2}{c_\text{m}}=\left(\f{1}{2}-\f{\Upsilon_2}{\Upsilon_1}\right)\f{\Sigma_2}{\Sigma_\text{m}}.
\end{equation}

The GI bifurcation is subcritical if the bracketed quantity on the left-hand side of equation~(\ref{wnl_amplitude}) is negative, because then nonlinear equilibria exist (for $k_1=0$, i.e.\ $k=k_\text{m}$) when $(Q^{-1})_2<0$, i.e.\ when $Q$ is (slightly) greater than~$1$. This condition for subcritical behaviour can be expressed as
\begin{equation}
  \f{3\Upsilon_3}{8\Upsilon_1}-\f{\Upsilon_2^2}{\Upsilon_1^2}<0
\end{equation}
or, equivalently, in terms of the higher logarithmic derivatives defined in equation~(\ref{higher_log}),
\begin{equation}
  \left[2-\Gamma^{(1)}\right]\left[5-3\Gamma^{(1)}\right]>\left[11-5\Gamma^{(1)}\right]\f{\Gamma^{(2)}}{\Gamma^{(1)}}-4\left[\f{\Gamma^{(2)}}{\Gamma^{(1)}}\right]^2+\f{\Gamma^{(3)}}{\Gamma^{(1)}}.
\end{equation}

In the 2D model considered by \citet{2022ApJ...934L..19D}, in which $P\propto\Sigma^\Gamma$, we have
\begin{equation}
  \f{\Upsilon_2}{\Upsilon_1}=-\f{2-\Gamma}{2},\qquad
  \f{\Upsilon_3}{\Upsilon_1}=\f{(3-\Gamma)(2-\Gamma)}{6},
\end{equation}
and so
\begin{equation}
  \f{3\Upsilon_3}{8\Upsilon_1}-\f{\Upsilon_2^2}{\Upsilon_1^2}=-\f{(2-\Gamma)(5-3\Gamma)}{16}.
\end{equation}
The bifurcation is subcritical for $\Gamma<5/3$ (or $\Gamma>2$), as stated above. If the correspondence $\Gamma=3-(2/\gamma)$ is made \citep{1972AnRFM...4..219H,1979SvA....23..153M}, as we also find in the affine model in the PSG limit (see equation~\ref{logderivs}), then the subcritical range translates into $\gamma<3/2$ (or $\gamma>2$).

Numerical evaluation of the $\Upsilon$ coefficients for the affine model with $\kappa=\nu$ and $\mathcal{W}=1.15$ suggests that the bifurcation is subcritical for $\gamma\lesssim1.50$ (or $\gamma\gtrsim2.00$), which is almost identical to the above translation of the 2D result and agrees with the findings of \citet{2025arXiv250111658B}. Importantly, the case $\gamma=1.4$, corresponding to a diatomic molecular gas in which rotational but not vibrational degrees of freedom are excited in equipartition, is subcritical.

The implication of this result is that warm molecular gas that behaves approximately adiabatically, or any gas that behaves approximately isothermally because of thermal relaxation to a temperature imposed by external radiation, is subject to subcritical GI behaviour in which some interesting nonlinear dynamics can occur in the linearly stable regime $Q>1$ when perturbations of finite amplitude are introduced. This could help to explain the self-sustaining processes at work in gravitational turbulence resulting from the GI, in which the average value of $Q$ is noticeably larger than~$1$.

\section{Summary and conclusion}
\label{s:conclusions}

In this paper we have reconsidered aspects of the theory of self-gravitating gaseous discs. The main results are as follows.

We have developed an expansion of the internal gravitational potential in powers of the aspect ratio of the disc (or of a structure within it) and separated the potential into near and far contributions. The near contribution corresponds to the potential of a slab with the same vertical structure as the disc has locally. The far contribution involves an integral over the area of the disc and can be expressed as a type of multipole expansion. The leading term involves only the surface density and corresponds to the usual potential considered for a 2D disc, while the next (quadrupolar) correction also involves the scaleheight.

By studying the hydrostatic vertical structure of a wide family of disc models, both analytically and numerically, we have shown that the near contribution to the self-gravitational energy can be written in an almost universal form in terms of the surface density and scaleheight. This has allowed us to develop an affine model of the dynamics of self-gravitating discs in which the scaleheight is free to evolve dynamically. The near gravitational energy acts as a gravitational pressure in the plane of the disc, adding significantly to the gas pressure and allowing us to define an enhanced effective sound speed and Toomre stability parameter $Q$ for self-gravitating discs such that the linear stability criterion is restored to $Q>1$.

This theory fairly accurately reproduces the onset of axisymmetric GI in 3D discs with resolved vertical structure. Among other things, this analysis shows that the critical radial wavelength is on the order of twenty times the scaleheight, helping to justify the validity of the affine model. The weakly nonlinear theory also typically exhibits subcritical behaviour, with equilibrium solutions of finite amplitude being found in the linearly stable regime $Q>1$ for adiabatic exponents less than $1.50$. This result is in accord with the fully 3D calculation of weakly nonlinear GI by \citet{2025arXiv250111658B} and is likely to play an important role in the understanding of the nonlinear outcome of GI in astrophysical discs, including the mechanism by which gravitational turbulence is self-sustaining.

\section*{Acknowledgements}

I am grateful to Joshua Brown, Hongping Deng, Henrik Latter and Roman Rafikov for discussions on this topic. I thank the referee, Giuseppe Lodato, for a very helpful report that led to improvements in the manuscript. This research was supported by STFC through grant ST/X001113/1.

\section*{Data Availability}

The data underlying this article will be shared on reasonable request to the author.

\bibliographystyle{mnras}
\bibliography{ms}

\appendix

\section{Asymptotic evaluation of the gravitational potential of a thin disc}
\label{s:appendix_potential}

The exact expression for the gravitational potential $\Phi_\text{d}$ due to the disc is given in equation~(\ref{phi_int_green}) and uses the 3D distance $\Delta_3$ between a general point $(x',y',z')$ in the disc and the point $(x,y,z)$ at which the potential is measured. We assume the disc is symmetric about the midplane.

Consider the evaluation of $\Phi_\text{d}$ at a particular horizontal location~$(x,y)$, where the scaleheight is~$H$. We assume that the characteristic lengthscale $L$ on which the disc varies in the horizontal directions is such that the aspect ratio $\epsilon=H/L$ is small $(\epsilon\ll1)$.

We exclude from the calculation any regions where $|z|\gg H$ (or $|z'|\gg H'$) because there is negligible mass at such distances from the midplane. In the isothermal model for vertical structure, the density is exponentially small for $|z|\gg H$, and in the polytropic models of finite index it is zero, so the error made in limiting $|z'|/H'$ to values of order unity is (at most) exponentially small in $\epsilon$.

Let $(r,\theta)$ be polar coordinates centred on the point $(x,y)$, such that $x'-x=r\cos\theta$, $y'-y=r\sin\theta$
and
\begin{equation}
  \Delta_2=r,\qquad
  \Delta_3=\sqrt{r^2+\left(z'-z\right)^2}.
\end{equation}
Note that $\Delta_3$ exceeds $\Delta_2$ by an amount that decreases from $\left|z'-z\right|$ to~$0$ as $r$ increases from $0$ to $\infty$.

If $\Delta_3$ is replaced by $\Delta_2$ in equation~(\ref{phi_int_green}), then the integral with respect to $z'$ can be carried out and we obtain the midplane potential of the razor-thin disc,
\begin{equation}
  -G\int\f{\Sigma'}{\Delta_2}\,\dd A',
\end{equation}
which is the leading term $\Phi_1$ (equation~\ref{phi1_integral}) in the expansion proposed in Section~\ref{s:expansion_potential}. As noted there, this integral is non-singular and can be evaluated without any regularization or smoothing.

We are therefore interested in the residual part of the integral,
\begin{equation}
  G\int\left(\f{1}{\Delta_2}-\f{1}{\Delta_3}\right)\,\rho'\,\dd V'.
\label{residual}
\end{equation}
To evaluate this positive integral, let us separate the region of integration into the interior and exterior of the cylinder $r=R$, where
$R$ is some intermediate distance such that $H\ll R\ll L$. (We will see below that $R$ does not need to be defined precisely.)

In the outer region $r>R$, we have $\left|z'-z\right|\ll R$, because regions with $\left|z'-z\right|\gg H$ are excluded from the calculation. Then we can expand
\begin{equation}
  \f{1}{\Delta_3}=\f{1}{\Delta_2}\left[1-\f{\left(z'-z\right)^2}{2\Delta_2^2}+O\left(\f{H^4}{R^4}\right)\right]
\end{equation}
and carry out the integration with respect to~$z'$, using the properties
\begin{equation}
  \int\rho'\,\dd z'=\Sigma',\quad
  \int\rho'z'\,\dd z'=0,\quad
  \int\rho'z'^2\,\dd z'=\Sigma'H'^2.
\end{equation}
Therefore the leading contribution to the residual integral~(\ref{residual}) from the outer region is
\begin{equation}
  G\int_{r>R}\f{\Sigma'(z^2+H'^2)}{2r^3}\,\dd A'.
\label{residual_outer}
\end{equation}
This integral is finite (assuming the disc to be finite) because the region of integration excludes the singularity at $r=0$. The error term is smaller by a factor of order $(H/R)^2$.

For the inner region $r<R$, we can use the property $r\ll L$ to expand the density in a Taylor series in $x'-x=r\cos\theta$ and $y'-y=r\sin\theta$. The leading term, $\rho(x,y,z')$, is independent of $x'$ and $y'$, and its contribution to the residual integral~(\ref{residual}) involves the horizontal integral
\begin{align}
  &\int_0^{2\pi}\int_0^R\left[\f{1}{r}-\f{1}{\sqrt{r^2+\left(z'-z\right)^2}}\right]r\,\dd r\,\dd\theta\nonumber\\
  &\qquad=2\pi\left[R-\sqrt{R^2+\left(z'-z\right)^2}+\left|z'-z\right|\right].
\end{align}
In the regions that are not excluded, this can be expanded as
\begin{equation}
  2\pi R\left[\f{\left|z'-z\right|}{R}-\f{\left(z'-z\right)^2}{2R^2}+O\left(\f{H^4}{R^4}\right)\right].
\end{equation}
The corresponding contributions to the residual integral~(\ref{residual}) from the inner region are, to the same order,
\begin{equation}
  2\pi G\int\rho\left(x,y,z'\right)\left|z'-z\right|\dd z'-\f{\pi G\Sigma\left(z^2+H^2\right)}{R}.
\label{residual_inner}
\end{equation}
Continuing the Taylor series for $\rho'$, we find that the first-order terms linear in $x'-x$ or $y'-y$ integrate to zero over the disc of radius $R$, because $\int_0^{2\pi}\cos\theta\,\dd\theta=0$ and similarly for $\sin\theta$. There are second-order terms that do not integrate to zero. These are smaller than the last term in equation~(\ref{residual_inner}) by a factor of order $R^2/L^2$ because they involve second derivatives of $\Sigma$ with respect to the horizontal coordinates.

When the contributions to the residual integral from inner and outer regions are combined, the result should not depend on the value of $R$, as long as it is in the intermediate range $H\ll R\ll L$. This implies some cancellation of terms. In fact, the potential $\Phi_2$ (equation~\ref{phi2_integral}) comes directly from the first term in equation~(\ref{residual_inner}). This is the `near' self-gravitational potential and clearly derives from the inner region.

The second term in equation~(\ref{residual_inner}) needs to be considered in conjunction with equation~(\ref{residual_outer}) from the outer region. Taken together, they produce a result that is independent of~$R$ in the limit under consideration and corresponds to the `Hadamard finite part' of the strongly singular integral $\Phi_3$ (equation~\ref{phi3_integral}). When written in polar coordinates centred on $(x,y)$, the integrand for $\Phi_3$ involves a double pole because of the factor $1/r^2$. The second term in equation~(\ref{residual_inner}) can be seen as coming from cutting out the non-integrable part of the integrand, leaving the finite part of the integral.

\section{Polytropic models}
\label{s:appendix_polytropic}

A polytropic model has
\begin{equation}
  p=K_3\rho^{1+1/n},
\label{polytropic_3d}
\end{equation}
where $n>0$ is the polytropic index (not necessarily an integer) and $K_3>0$ is the (3D) polytropic constant. A vertically polytropic disc is one in which $K_3$ is independent of $z$ for a particular choice of $n$. The limits $n\to0$ and $n\to\infty$ correspond to homogeneous ($\rho=\cst$) and isothermal ($p\propto\rho$) models, respectively.

A polytrope occurs naturally if the disc is composed of a perfect gas of adiabatic exponent $\gamma$ and its specific entropy is independent of $z$. In this case of an adiabatically stratified disc, $\gamma=1+1/n$, and we make use of this correspondence.
The specific enthalpy
\begin{equation}
  h=(n+1)K_3\rho^{1/n}
\end{equation}
satisfies $\rho\,\dd h=\dd p$. 

Let us introduce dimensionless pressure and potential functions $F_p$ and $F_\Phi$ such that
\begin{equation}
  p=\f{P}{H}F_p(\zeta),\qquad
  \Phi=\pi G\Sigma HF_\Phi(\zeta).
\end{equation}
Then the definition of $P$ (equation~\ref{sigma_h_p}) implies
\begin{equation}
  \int F_p(\zeta)\,\dd\zeta=1,
\label{normalization_p}
\end{equation}
while hydrostatic balance (equation~\ref{hydrostatic}) becomes
\begin{equation}
  -\f{1}{F_\rho}\f{\dd F_p}{\dd\zeta}=(1-s)\zeta+\f{s}{\mathcal{W}}\f{\dd F_\Phi}{\dd\zeta}
\label{hydrostatic_polytropic}
\end{equation}
(in which we also used equations~\ref{w} and~\ref{s}) and Poisson's equation~(\ref{poisson}) becomes
\begin{equation}
  \f{\dd^2F_\Phi}{\dd\zeta^2}=4F_\rho,
\end{equation}
with solution
\begin{equation}
  F_\Phi(\zeta)=2\int\left|\zeta-\zeta'\right|F_\rho(\zeta')\,\dd\zeta'.
\end{equation}
In the regions above and below the disc, where $F_\rho$ is zero (or negligible), we have $F_\Phi=2|\zeta|$, corresponding to a uniform gravitational field. For example, above the disc, we can replace $|\zeta-\zeta'|$ in this integral with $(\zeta-\zeta')$; the $\zeta'$ term integrates to zero by symmetry and then the $\zeta$ factor can be taken out of the integral.

For an adiabatically stratified disc, the dimensionless entropy
\begin{equation}
  \mathcal{S}=\f{F_p^n}{F_\rho^{n+1}}
\end{equation}
is independent of $\zeta$. Furthermore, the 3D polytropic relation~(\ref{polytropic_3d}) implies
\begin{equation}
  \f{P}{H}=K_2\left(\f{\Sigma}{H}\right)^{1+1/n},
\label{polytropic_2d}
\end{equation}
where $K_2$, $K_3$ and $\mathcal{S}$ are related by
\begin{equation}
  \mathcal{S}=\left(\f{K_3}{K_2}\right)^n.
\end{equation}
Equation~(\ref{polytropic_2d}) looks like a 2D polytropic relation connecting the surface density, scaleheight and vertically integrated pressure. However, consider a one-parameter family of vertically polytropic solutions with the same value of $K_3$. The family can be parametrized by $0<s<1$. The value of $K_2$ is not generally constant within this family, because $\mathcal{S}$ depends on $s$. However, we will see below that the variation of $\mathcal{S}$ with $s$ is extremely limited.

The dimensionless enthalpy
\begin{equation}
  F_h=(n+1)\left(\mathcal{S}F_\rho\right)^{1/n}=(n+1)\f{F_p}{F_\rho}
\end{equation}
satisfies $F_\rho\,\dd F_h=\dd F_p$ when $\mathcal{S}$ is constant, and the left-hand side of equation~(\ref{hydrostatic_polytropic}) becomes $-\dd F_h/\dd\zeta$. Eliminating $F_\Phi$ between the hydrostatic and Poisson equations gives the Lane--Enden-type equation
\begin{equation}
  \f{\dd^2F_h}{\dd\zeta^2}+\f{4s}{\mathcal{W}\mathcal{S}}\left(\f{F_h}{n+1}\right)^n+(1-s)=0.
\label{ode_f_h}
\end{equation}
This is to be solved in conjunction with the boundary conditions $F_h(
\pm\zeta_\text{s})=0$, where the dimensionless surface height $\zeta_\text{s}$ is to be determined, and with the normalization conditions
\begin{equation}
  \int\left(\f{F_h}{n+1}\right)^n\,\dd\zeta=\int\left(\f{F_h}{n+1}\right)^n\zeta^2\,\dd\zeta=\mathcal{S},
\label{normalization_polytropic}
\end{equation}
deriving from equation~(\ref{normalization}). Equation~(\ref{normalization_p}) implies further that
\begin{equation}
  \int\left(\f{F_h}{n+1}\right)^{n+1}\,\dd\zeta=\mathcal{S}.
\end{equation}
This is not an independent constraint as it can be derived from equations (\ref{ode_f_h}) and (\ref{normalization_polytropic}) after an integration by parts.

Equation~(\ref{ode_f_h}) does not involve $\zeta$ explicitly and has the first integral
\begin{equation}
  \f{1}{2}\left(\f{\dd F_h}{\dd\zeta}\right)^2+\f{4s}{\mathcal{W}\mathcal{S}}\left(\f{F_h}{n+1}\right)^{n+1}+(1-s)F_h=\text{constant}.
\end{equation}
At $\zeta=0$, where $F_h=F_{h0}$ takes its maximum value and $\dd F_h/\dd\zeta=0$, the value of the constant can be found to be
\begin{equation}
  \f{4s}{\mathcal{W}\mathcal{S}}\left(\f{F_{h0}}{n+1}\right)^{n+1}+(1-s)F_{h0}.
\end{equation}
Therefore, given the value of $F_{h0}$, we know $\dd F_h/\dd\zeta$ in terms of $F_h$ and the problem is reduced to quadrature.

For numerical purposes, however, we solve the problem by the following method. The dimensional differential equation for $h(z)$ is
\begin{equation}
  \f{\dd^2h}{\dd z^2}+4\pi G\rho+\nu^2=0,\qquad
  \rho=\left[\f{h}{(n+1)K_3}\right]^n.
\label{ode_h}
\end{equation}
At the midplane $z=0$, we have $\dd h/\dd z=0$. At the free upper surface $z=Z$, we have $h=0$ and $-\dd h/\dd z=\nu^2Z+2\pi G\Sigma$, which is the surface gravity.

To make the problem dimensionless, we introduce the units
\begin{equation}
  \rho_*=\f{\nu^2}{4\pi G},\qquad
  c_*^2=K_3\rho_*^{1/n},\qquad
  z_*=\f{c_*}{\nu}
\end{equation}
and define dimensionless variables (with tildes) by
\begin{equation}
  h=c_*^2\tilde h(\tilde z),\qquad
  \rho=\rho_*\tilde\rho(\tilde z),\qquad
  z=z_*\tilde z,\label{nondim}
\end{equation}
as well as
\begin{equation}
  Z=z_*\tilde Z,\qquad
  \Sigma=\rho_*z_*\tilde\Sigma,
\end{equation}
which then satisfy
\begin{equation}
  \f{\dd^2\tilde h}{\dd\tilde z^2}+\tilde\rho+1=0,\qquad
  \tilde\rho=\left(\f{\tilde h}{n+1}\right)^n.
\end{equation}
At the midplane $\tilde z=0$, we have $\dd\tilde h/\dd\tilde z=0$. At the free upper surface $\tilde z=\tilde Z$, we have $\tilde h=0$ and $-\dd\tilde h/\dd\tilde z=\tilde Z+\tilde\Sigma/2$.

Guessing the value of $\tilde Z$, we integrate the differential equation~(\ref{ode_h}) from $\tilde z=\tilde Z$ to $\tilde z=0$ and adjust $\tilde Z$ by Newton iteration to match the boundary condition that $\dd\tilde h/\dd\tilde z=0$ at $\tilde z=0$. We regard $\tilde\Sigma$ (as well as $\nu^2$ and $K_3$) as a known quantity. We thereby obtain a family of solutions (for each $n$) parametrized by $\tilde\Sigma$. For each member we can then evaluate $H$, $P$, $W$, etc., as follows:
\begin{align}
  &\Sigma=2\int_0^Z\rho\,\dd z=2\rho_*z_*\int_0^{\tilde Z}\tilde\rho\,\dd\tilde z,\\
  &\Sigma H^2=2\int_0^Z\rho z^2\,\dd z=2\rho_*z_*^3\int_0^{\tilde Z}\tilde\rho\tilde z^2\,\dd\tilde z,\\
  &P=2\int_0^ZK_3\rho^{1+1/n}\,\dd z=2\nu^2\rho_*z_*^3\int_0^{\tilde Z}\tilde\rho^{1+1/n}\,\dd\tilde z,\\
  &W=2\int_0^Z\rho z\f{\dd\Phi}{\dd z}\,\dd z=-2\int_0^Z\rho z\left(\f{\dd h}{\dd z}+\nu^2z\right)\,\dd z\nonumber\\
  &\qquad=-2\nu^2\rho_*z_*^3\int_0^Z\tilde\rho\tilde z\left(\f{\dd\tilde h}{\dd\tilde z}+\tilde z\right)\,\dd\tilde z.
\end{align}
We can then check the virial relation $P=W+\nu^2\Sigma H^2$ and deduce the dimensionless quantities $s=W/P$, $\mathcal{W}=W/\pi G\Sigma^2H$ and 
\begin{equation}
  \mathcal{S}=K_3^n\left(\f{P}{H}\right)^{-n}\left(\f{\Sigma}{H}\right)^{n+1}.
\end{equation}
(The dimensional factors cancel when calculating these.)

The problem has two parameters, $n$ and $s$. In the following subsections, we first discuss analytical solutions that can be found in special cases and then describe numerical results for other parameter values.

\subsection{Non-self-gravitating case}

In the NSG limit ($s=0$), equation~(\ref{ode_f_h}) is linear and the solution is \citep[cf.][]{2014MNRAS.445.2621O}
\begin{equation}
  F_h=\f{1}{2}\left(\zeta_\text{s}^2-\zeta^2\right),\qquad\zeta_\text{s}=\sqrt{2n+3},
\end{equation}
for $|\zeta|<\zeta_\text{s}$ (and zero otherwise). We then find
\begin{align}
  &\mathcal{W}=\f{16\sqrt{2n+3}\,[\Gamma(2n+2)]^4}{\Gamma(4n+5)[\Gamma(n+1)]^4},\label{c_nsg}\\
  &\mathcal{S}=\sqrt{(2n+3)\pi}\left(\f{2n+3}{2n+2}\right)^n\f{\Gamma(n+1)}{\Gamma\left(n+\f{3}{2}\right)}.\label{fs_nsg}
\end{align}
The variation of these quantities is discussed in the main text.

\subsection{Purely self-gravitating case}

In the PSG limit ($s=1$) the non-dimensionalization~(\ref{nondim}) can still be used, retaining $\nu$ as a dummy parameter and dropping the terms due to external gravity, so that the differential equation simplifies to
\begin{equation}
  \f{\dd^2\tilde h}{\dd\tilde z^2}+\tilde\rho=0,\qquad
  \tilde\rho=\left(\f{\tilde h}{n+1}\right)^n,\label{ode_psg}
\end{equation}
and the upper boundary condition to $-\dd\tilde h/\dd\tilde z=\tilde\Sigma/2$.
The first integral of the differential equation implies that
\begin{equation}
  \f{\dd\tilde h}{\dd\tilde z}=-\sqrt{2\left[\left(\f{\tilde h_0}{n+1}\right)^{n+1}-\left(\f{\tilde h}{n+1}\right)^{n+1}\right]}
\end{equation}
above the midplane, where $\tilde h_0=\tilde h(0)$. (From here it is possible to express $\tilde z$ in terms of $\tilde h$ using the incomplete beta function.) Using the replacement
\begin{equation}
  \dd\tilde z=\left(\f{\dd\tilde h}{\dd\tilde z}\right)^{-1}\tilde h_0\,\dd x,
\end{equation}
where $x=\tilde h/\tilde h_0$, we can convert integrals such as
\begin{equation}
  \tilde Z=\int_0^{\tilde Z}\,\dd\tilde z,\qquad
  \tilde\Sigma=2\int_0^{\tilde Z}\,\tilde\rho\,\dd\tilde z,\qquad
  \text{etc.,}
\end{equation}
into integrals with respect to $x$ (from $0$ to $1$). To evaluate $W$ and $H$ we make use of the simplified virial relation $P=W$ and the identity
\begin{equation}
  2\int_0^{\tilde Z}\tilde\rho\tilde z^2\,\dd\tilde z=\tilde\Sigma\tilde Z^2-4\int_0^{\tilde Z}\tilde h\,\dd\tilde z,
\end{equation}
which follows from the equation~(\ref{ode_psg}) after multiplication by $\tilde z^2$ and integration by parts.

The required dimensionless integrals are
\begin{align}
  &I_0=\int_0^1\f{\dd x}{\sqrt{1-x^{n+1}}}=\f{\sqrt{\pi}\,\Gamma\left(1+\f{1}{n+1}\right)}{\Gamma\left(\f{1}{2}+\f{1}{n+1}\right)},\\
  &I_1=\int_0^1\f{x\,\dd x}{\sqrt{1-x^{n+1}}}=\f{\sqrt{\pi}\,\Gamma\left(1+\f{2}{n+1}\right)}{2\Gamma\left(\f{1}{2}+\f{2}{n+1}\right)},\\
  &I_2=\int_0^1\f{x^n\,\dd x}{\sqrt{1-x^{n+1}}}=\f{2}{n+1},\\
  &I_3=\int_0^1\f{x^{n+1}\,\dd x}{\sqrt{1-x^{n+1}}}=\f{2I_0}{n+3}.
\end{align}
($I_0$ decreases monotonically from $2$ at $n=0$ to $1$ as $n\to\infty$, while $I_1$ decreases from $4/3$ to $1/2$, $I_2$ from $2$ to $0$ and $I_3$ from $4/3$ to $0$.)
We then find
\begin{align}
  \mathcal{W}&=\f{2I_0}{(n+3)\sqrt{I_0^2-2I_1}},\label{c_psg}\\
  \mathcal{S}&=\f{4}{(n+1)^{n+1}\sqrt{I_0^2-2I_1}}\left(\f{n+3}{I_0}\right)^n=2\left[\f{n+3}{(n+1)I_0}\right]^{n+1}\mathcal{W}.\label{fs_psg}
\end{align}
The variation of $\mathcal{W}$ and $\mathcal{S}$ with $n$ is described in the main text.

\subsection{Polytrope of index $n=1$}
\label{s:polytrope_n=1}

In the case $n=1$, the Lane--Emden-type equation~(\ref{ode_f_h}) is linear. The relevant solution can be written as
\begin{equation}
  F_h=\f{1-s}{k^2}\left[\f{\cos(k\zeta)}{\cos(k\zeta_\text{s})}-1\right],
\end{equation}
valid for $|\zeta|<\zeta_\text{s}$, with
$k^2=2s/\mathcal{W}\mathcal{S}$.
The one-parameter family of solutions can be parametrized by $\theta=k\zeta_\text{s}$, which is related to $s$ by
\begin{equation}
  s=\f{30(t-\theta)+\theta\left(8\theta^2-24\theta t+6t^2\right)}{6\left[\theta t^2-3(t-\theta)\right]},\label{s_n=1}
\end{equation}
where $t=\tan\theta$. As $\theta$ increases from $0$ to $\pi/2$, $s$ increases from $0$ to~$1$.

Using the definition of $\mathcal{W}$ and the normalizing integrals, we find
\begin{align}
  &\mathcal{W}^2=\f{4s^2\left[\theta(6-\theta^2)-3(2-\theta^2)t\right]}{3(1-s)^2(t-\theta)^3},\\
  &k=\f{\mathcal{W}(1-s)(t-\theta)}{2s},\\
  &\mathcal{S}=\f{2s}{\mathcal{W}k^2}=\f{(2s)^3}{\mathcal{W}^3(1-s)^2(t-\theta)^2}.
\end{align}

This implies that $\mathcal{W}$ decreases monotonically from
\begin{equation}
  \sqrt{\f{324}{245}}\approx1.1500\qquad\text{to}\qquad
  \f{\pi}{2\sqrt{\pi^2-8}}\approx1.1488
\end{equation}
as $s$ increases from $0$ to $1$. Meanwhile $\mathcal{S}$ decreases monotonically from
\begin{equation}
  \f{5\sqrt{5}}{3}\approx3.7268\qquad\text{to}\qquad
  \f{16}{\pi\sqrt{\pi^2-8}}\approx3.7247
\end{equation}
and $\zeta_\text{s}\theta/k$ increases monotonically from
\begin{equation}
  \sqrt{5}\approx2.2361\qquad\text{to}\qquad
  \f{\pi}{\sqrt{\pi^2-8}}\approx2.2976.
\end{equation}

There are simple expressions for $F_\rho$ in two limits. in the NSG limit $s=0$ we have (for $|\zeta|<\sqrt{5}$)
\begin{equation}
  F_\rho(\zeta)=\f{3}{4\sqrt{5}}\left(1-\f{\zeta^2}{5}\right),
\label{density_n=1_nsg}
\end{equation}
and in the PSG limit $s=1$ we have (for $|\zeta|<\pi/2k$)
\begin{equation}
  F_\rho(\zeta)=\f{k}{2}\cos(k\zeta),\qquad k=\f{\sqrt{\pi^2-8}}{2}.
\label{density_n=1_psg}
\end{equation}

\subsection{Homogeneous case}

The homogeneous case ($\rho$ independent of $z$) corresponds to the limit $n\to0$ of the polytropic model.  Independent of $s$, we have
\begin{equation}
  F_\rho(\zeta)=\begin{cases}
    \f{1}{2\sqrt{3}},&|\zeta|<\sqrt{3},\\
    0,&|\zeta|>\sqrt{3},
  \end{cases}
\label{density_homogeneous}
\end{equation}
\begin{align}
  &\mathcal{W}=\f{2}{\sqrt{3}}\approx1.1547,\\
  &\mathcal{S}=2\sqrt{3}\approx3.4641.
\end{align}

\subsection{Isothermal case}

The isothermal case ($p\propto\rho$) corresponds to the limit $n\to\infty$ of the polytropic model. Unlike the case of finite $n$, the isothermal disc does not have a definite upper surface at which the density and pressure vanish. Although the solutions formally extend to arbitrarily large $|\zeta|$, there is negligible mass where $|\zeta|\gg1$. Well-known analytical solutions are possible in two limits:

In the NSG limit $s=0$ we have
\begin{align}
  &F_\rho(\zeta)=\f{1}{\sqrt{2\pi}}\exp\left(-\f{\zeta^2}{2}\right),\label{density_isothermal_nsg}
\\
  &\mathcal{W}=\f{2}{\sqrt{\pi}}\approx1.1284,\\
  &\mathcal{S}=\sqrt{2\pi\ee}\approx4.1327.
\end{align}

In the PSG limit $s=1$ we have
\begin{align}
  &F_\rho(\zeta)=\f{\pi}{4\sqrt{3}}\sech^2\left(\f{\pi\zeta}{2\sqrt{3}}\right),\label{density_isothermal_psg}\\
  &\mathcal{W}=\f{2\sqrt{3}}{\pi}\approx1.1027,\\
  &\mathcal{S}=\f{\sqrt{3}\,\ee^2}{\pi}\approx4.0738.
\end{align}

\section{Analytical results for marginal gravitational stability}
\label{s:appendix_marginal}

In this appendix we derive exact results for marginal gravitational stability in two special cases that can be treated analytically, and compare these results with those of the affine model. Our treatments are closely related to those of \citet{1965MNRAS.130...97G} but differ in detail because we allow for differential rotation and a central potential.

\subsection{Polytrope of index $n=1$}
\label{s:appendix_n=1}

The $n=1$ polytrope has pressure $p=K_3\rho^2$ and specific enthalpy $h=2K_3\rho$ such that $\dd p=\rho\,\dd h$. The equilibrium satisfies the Lane--Emden-type equation
\begin{equation}
  \f{\dd^2h}{\dd z^2}+k_0^2h+\nu^2=0,
\label{lane-emden_n=1}
\end{equation}
equivalent to equation~(\ref{ode_h}), where $k_0=(2\pi G/K_3)^{1/2}$
is a characteristic wavenumber. The relevant solution, symmetric about the midplane and having free surfaces at $z=\pm Z$, is, for $|z|<Z$,
\begin{align}
  &h=\f{\nu^2}{k_0^2}\left(\f{\cos k_0z}{\cos k_0Z}-1\right),\\
  &\rho=\f{\nu^2}{4\pi G}\left(\f{\cos k_0z}{\cos k_0Z}-1\right),\\
  &\Phi=-\f{\nu^2}{k_0^2}\f{\cos k_0z}{\cos k_0Z}-\f{1}{2}\nu^2z^2+\text{constant},
\end{align}
and the surface density is
\begin{equation}
  \Sigma=\f{\nu^2}{2\pi Gk_0}(t-\theta),
\end{equation}
where $t=\tan\theta$ and $\theta=k_0Z$ is the same parameter as in Section~\ref{s:polytrope_n=1}. Since $t-\theta$ increases monotonically from $0$ to $+\infty$ as $\theta$ increases from $0$ to $\pi/2$, so there is a unique $\theta$ in this range for any positive $\Sigma$. The degree of self-gravity, $s$, is given by equation~(\ref{s_n=1}) and increases monotonically from $0$ to $1$ as $\theta$ increases from $0$ to $\pi/2$. We will also need
\begin{equation}
  \f{H}{Z}=\left[\f{2\theta}{3(t-\theta)}+1-\f{2}{\theta^2}\right]^{1/2},
\end{equation}
which decreases monotonically from $1/\sqrt{5}=0.4472$ to $\sqrt{1-(8/\pi^2)}=0.4352$ as $\theta$ increases from $0$ to $\pi/2$.

We now identify a marginally stable equilibrium by looking for a solution of the linearized equations proportional to $\ee^{\ii kx}$. The perturbed solution represents a nearby $x$-dependent equilibrium state that is reached via a time-independent meridional displacement $(\xi_x,\xi_z)$ that preserves the angular momentum of each fluid element. (This constraint is equivalent to assuming an isovortical perturbation.) Taking into account the associated Eulerian velocity perturbation $\delta u_y$, we obtain the meridional force balances
\begin{align}
  &\kappa^2\xi_x=-\ii k\left(\delta\Phi+\delta h\right),\\
  &0=-\f{\dd}{\dd z}\left(\delta\Phi+\delta h\right).
\end{align}
Perturbations of enthalpy and density are related by
\begin{equation}
  \f{\delta h}{2K_3}=\delta\rho=-\rho\ii k\xi_x-\f{\dd}{\dd z}(\rho\xi_z),
\label{delta_rho_n=1}
\end{equation}
and Poisson's equation gives
\begin{equation}
  \left(-k^2+\f{\dd^2}{\dd z^2}\right)\delta\Phi=k_0^2\,\delta h.
\end{equation}
Thus
\begin{equation}
  \left(k_0^2-k^2+\f{\dd^2}{\dd z^2}\right)\delta\Phi=k_0^2A,
\label{poisson_lin_n=1}
\end{equation}
where
\begin{equation}
  A=\delta\Phi+\delta h=-\f{\kappa^2\xi_x}{\ii k}=\cst.
\label{a_n=1}\end{equation}
The relevant critical solution of equation~(\ref{poisson_lin_n=1}) turns out to be a symmetric mode with $k<k_0$. With $k_z=(k_0^2-k^2)^{1/2}$
we have
\begin{equation}
  \delta\Phi=\f{k_0^2}{k_z^2}A+B\cos k_zz,
\end{equation}
and so
\begin{equation}
  \delta h=-\f{k^2}{k_z^2}A-B\cos k_zz.
\label{delta_h_n=1}
\end{equation}
At the surface $z=Z$, $\delta\Phi$ and its first derivative must match continuously with a decaying solution $\propto\ee^{-|k|z}$ in the vacuum region $z>Z$. Thus
\begin{equation}
  -Bk_z\sin k_zZ=-|k|\left(\f{k_0^2}{k_z^2}A+B\cos k_zZ\right).
\label{ab1}
\end{equation}

The unknown $\xi_z$ can be eliminated from equation~(\ref{delta_rho_n=1}) by integrating vertically through the disc, giving the surface density perturbation
\begin{equation}
  \f{1}{2K_3}\int\,\delta h\,\dd z=\delta\Sigma=-\ii k\int\rho\xi_x\,\dd z=-\f{k^2\Sigma A}{\kappa^2}.
\end{equation}
Using the expression~(\ref{delta_h_n=1}) for $\delta h$ and multiplying by $-K_3$ leads to
\begin{equation}
  \f{k^2}{k_z^2}AZ+\f{B}{k_z}\sin k_zZ=\f{\nu^2}{\kappa^2}\f{k^2}{k_0^3}A(t-\theta).
\label{ab2}
\end{equation}
Equations (\ref{ab1}) and (\ref{ab2}) for $A$ and $B$ are compatible if
\begin{equation}
  \f{|k|k_z}{k_0^3}\left[\f{\nu^2}{\kappa^2}\f{k_z^2}{k_0^2}(t-\theta)-\theta\right]\left(k_z-|k|\cot k_zZ\right)=1.
\label{marginal_n=1}
\end{equation}
For a given value of $\nu^2/\kappa^2$, this equation has no real positive roots for $|k|<k_0$ if $\theta$ (and therefore $\Sigma$) is smaller than a critical value, but two such roots if $\theta$ is sufficiently large. The critical value, at which the two roots coalesce, corresponds to marginal gravitational stability of the disc.

When $\nu^2/\kappa^2=1$, marginal stability is found at $\theta=k_0Z=1.4311$ and $|k|Z=0.5316$, implying $s=0.8390$ and $|k|H=0.2326$. These values are not very sensitive to the value of $\nu^2/\kappa^2$ within the range expected for thin discs (see Section~\ref{s:total}). For example, when $\nu^2/\kappa^2=0.9$, marginal stability occurs at $\theta=1.4424$, $s=0.8517$ and $|k|H=0.2348$.

It is also notable that $\pi G\Sigma k/\kappa^2=1.0551$ for $\nu^2/\kappa^2=1$ (and is only very slightly dependent on $\nu^2/\kappa^2$). Therefore the critical wavenumber is very close to the value $\kappa^2/\pi G\Sigma$ predicted by the simplest 2D theory but also by the affine model neglecting quadrupolar gravity.

However, in order to compare with the PSG (and uniformly rotating) disc considered by \citet{1965MNRAS.130...97G}, we need to take the limits $\theta\to\pi/2$ (i.e.\ $s=1$) and $\nu^2/\kappa^2\to0$. A relevant measure of self-gravity in this limit is $\pi G\Sigma k_0/\kappa^2$.
We need to rewrite the marginal condition as
\begin{equation}
  \f{|k|k_z}{k_0^3}\left(\f{k_z^2}{k_0^2}\f{2\pi G\Sigma k_0}{\kappa^2}-\f{\pi}{2}\right)\left(k_z-|k|\cot k_zZ\right)=1.
\end{equation}
The numerical solution is $|k|Z=0.5955$, $|k|H=0.2592$ and $\pi G\Sigma k_0/\kappa^2=2.8072$. \citet{1965MNRAS.130...97G} quote (for the case of no halo pressure) $\pi G\bar\rho/\kappa^2=1.11$ (note that $\kappa^2=4\Omega^2$ for their uniformly rotating disc) and $|k|T=0.97$, where the mean density and `thickness' are related to our variables by $\bar\rho=(\pi^2/16)(\Sigma/Z)$ and $T=(16/\pi^2)Z$. So their results translate into $|k|Z=0.60$ and $\pi G\Sigma k_0/\kappa^2=(1.11)(8/\pi)=2.83$, which agree with our numerical solution to two significant figures.

For comparison with the affine theory, let us rearrange the marginal condition (\ref{marginal_n=1}) in the form
\begin{equation}
  \f{\kappa^2}{\nu^2}=f\left(\f{|k|}{k_0}\right)
\end{equation}
and expand the function $f(x)=f_1x+f_2x^2+f_3x^3+f_4x^4+\cdots$ in a Taylor series. The first four terms have the form
\begin{align}
  &f_1=t-\theta,\\
  &f_2=-(t-\theta)\left(\f{1}{t}+\theta\right),\\
  &f_3=(t-\theta)\left[\f{2\theta}{t}+(\theta^2-2)\right],\\
  &f_4=-\f{1}{2}(t-\theta)\left[\f{3\theta}{t^2}+\f{3(2\theta^2-1)}{t}+\theta(2\theta^2-5)\right].
\end{align}
If we truncate this expansion at various terms and use it to estimate marginal stability (through occurrence of a double root for $x$) in the case $\nu^2/\kappa^2=1$, we find:
\begin{itemize}
\item $\theta\approx1.4421$ ($s=0.8513$, $x=0.3182$) if truncated at $f_2$
\item $\theta\approx1.4300$ ($s=0.8378$, $x=0.3803$) if truncated at $f_3$
\item $\theta\approx1.4306$ ($s=0.8385$, $x=0.3754$) if truncated at $f_4$
\end{itemize}
These results show rapid convergence towards the correct value of $\theta=1.4311$ ($s=0.8390$, $x=0.3715$) as more terms are included.

In the affine model, the $n=1$ polytropic sequence has
\begin{align}
  &H=\f{K_2^{1/2}}{(\mathcal{W}\pi G)^{1/2}}s^{1/2},\\
  &\Sigma=\f{K_2^{1/2}\nu^2}{(\mathcal{W}\pi G)^{3/2}}\f{s^{3/2}}{1-s},\\
  &\Pi=\f{K_2^{3/2}\nu^4}{(\mathcal{W}\pi G)^{5/2}}\f{s^{5/2}(1+s)}{(1-s)^2},\\
  &c^2=\f{K_2\nu^2}{\mathcal{W}\pi G}\f{s(5+6s-3s^2)}{(1-s)(3-s)}.
\end{align}
The equivalent coefficients of the quartic marginal stability condition~(\ref{marginal_quartic}) resulting from the affine model including quadrupolar gravity are
\begin{align}
  f_1&\approx\f{2\pi G\Sigma k_0}{\nu^2}=\left(\f{2}{C}\right)^{3/2}\mathcal{S}^{-1/2}\f{s^{3/2}}{1-s},\\
  f_2&\approx-\f{c^2k_0^2}{\nu^2}\approx-\f{2}{C\mathcal{S}}\f{s(5+6s-3s^2)}{(1-s)(3-s)},\\
  f_3&\approx\left(\f{5-3s}{3-s}\right)\f{2\pi G\Sigma k_0^3H^2}{\nu^2}\approx\left(\f{2}{C}\right)^{5/2}\mathcal{S}^{-3/2}\left(\f{5-3s}{3-s}\right)\f{s^{5/2}}{1-s},\\
  f_4&\approx\f{(2\pi G\Sigma)^2(k_0H)^4}{\nu^2\left(\f{2P}{\Sigma}+\nu^2H^2\right)}\approx\left(\f{2}{C}\right)^4\mathcal{S}^{-2}\f{s^4}{(1-s)(3-s)}.
\end{align}
These coefficients, evaluated for the affine polytropic sequence, with $\mathcal{W}\approx1.149$ and $K_3/K_2=\mathcal{S}\approx3.725$, agree almost exactly with the exact ones above, in the case of $f_1$, $f_2$ and $f_3$. So the affine model is accurate when the marginal condition is truncated at cubic order.  However, the expression for $f_4$ is completely inaccurate (as for the homogeneous model) and should not be used to estimate marginal stability.

\subsection{Equilibrium and marginal stability in a homogeneous incompressible disc}
\label{s:appendix_homogeneous}

The equilibrium solution in the interior $|z|<Z$ of a disc of uniform density $\rho$ and semithickness $Z$ has $\Phi=2\pi G\rho\left(Z^2+z^2\right)$ and $p=\f{1}{2}\rho\left(4\pi G\rho+\nu^2\right)\left(Z^2-z^2\right)$. The integrated quantities are $\Sigma=2\rho Z$, $\Sigma H^2=\f{2}{3}\rho Z^3$ (giving $H=Z/\sqrt{3}$), $P=\f{2}{3}\left(4\pi G\rho+\nu^2\right)\rho Z^3$ and $W=\f{8}{3}\pi G\rho^2Z^3$. The virial relation~(\ref{virial}) is satisfied and the degree of self-gravity is $s=4\pi G\rho/\left(4\pi G\rho+\nu^2\right)$.

At marginal stability for an incompressible fluid we have, for $|z|<Z$, the linearized equation of motion
\begin{align}
  &\kappa^2\xi_x=-\ii k\left(\Phi'+\f{p'}{\rho}\right),\\
  &0=-\p_z\left(\Phi'+\f{p'}{\rho}\right),
\end{align}
together with Poisson's equation
\begin{equation}
  \p_z^2\Phi'-k^2\Phi'=0
\label{poisson_homogeneous}
\end{equation}
and the incompressibility condition
\begin{equation}
  \ii k\xi_x+\p_z\xi_z=0.
\end{equation}
It follows that $\xi_x=\text{constant}$ and $\xi_z\propto z$, as assumed in the affine model. Indeed $\xi_z=-\ii kz\xi_x$. The relevant solution of equation~(\ref{poisson_homogeneous}), symmetric about the midplane, has $\Phi'=A\cosh kz$ for some~$A$. 
In the exterior $|z|>Z$ we also have equation~(\ref{poisson_homogeneous}), but here the relevant solution, decaying at infinity, has $\Phi'=B\,\exp\left(-|k||z|\right)$ for some~$B$. At $z=Z$, $\Phi'$ is continuous, while $\p_z\Phi'$ increases by $4\pi G\rho\xi_z$ owing to the Eulerian density perturbation $\xi_z\rho\,\delta(z-Z)$. Thus
\begin{align}
  &B\,\ee^{-|k|Z}-A\cosh kZ=0,\\
  &-|k|B\,\ee^{-|k|Z}-kA\sinh kZ=-4\pi G\rho\,\ii kZ\xi_x.
\end{align}
A further condition is that the Lagrangian pressure perturbation $p'+\xi_z\p_zp$ vanishes at the surfaces. Hence
\begin{equation}
  -\rho\f{\kappa^2\xi_x}{\ii k}-\rho A\cosh kZ+\ii kZ\xi_x\rho\left(4\pi G\rho+\nu^2\right)Z=0.
\end{equation}
Eliminating $A$ and $B$ between the last three equations leads to the marginal stability condition
\begin{equation}
  \left(1+|\tanh kZ|\right)|k|Z\left(4\pi G\rho+\nu^2+\f{\kappa^2}{k^2Z^2}\right)=4\pi G\rho.
\label{marginal_incompressible}
\end{equation}

In the case $\nu^2=\kappa^2$ we have a critical equilibrium at $4\pi G\rho/\kappa^2=7.6169$ and $kZ=0.2775$, corresponding to $s=0.8839$ and $kH=0.1602$.

In the case $\nu^2=0$ we have a critical equilibrium at $4\pi G\rho/\kappa^2=7.0255$ and $kZ=0.3035$ (in agreement with \citealt{1965MNRAS.130...97G}), corresponding to $kH=0.1752$.

For comparison with the affine theory, let us rearrange the marginal condition (\ref{marginal_incompressible}) in the form
\begin{equation}
  \f{\kappa^2}{\nu^2}=f(|k|Z),
\end{equation}
with
\begin{equation}
  f(x)=\f{\tilde\rho x}{1+\tanh x}-(\tilde\rho+1)x^2
\end{equation}
and $\tilde\rho=4\pi G\rho/\nu^2$. This has the Taylor series
\begin{equation}
  f(x)=\tilde\rho x-(2\tilde\rho+1)x^2+\tilde\rho x^3-\f{2}{3}\tilde\rho x^4+\cdots.
\end{equation}
If we truncate this expansion at various terms and use it to estimate marginal stability (through occurrence of a double root for $x$) in the case $\nu^2/\kappa^2=1$, we find:
\begin{itemize}
\item $\tilde\rho\approx8.472$ ($s=0.8944$, $x=0.2361$) if truncated at $f_2$
\item $\tilde\rho\approx7.414$ ($s=0.8811$, $x=0.2956$) if truncated at $f_3$
\item $\tilde\rho\approx7.643$ ($s=0.8843$, $x=0.2748$) if truncated at $f_4$
\end{itemize}
These results show convergence towards the correct value of $\tilde\rho\approx7.617$ ($s=0.8839$, $x=0.2775$) as more terms are included.

The marginal condition~(\ref{marginal_quartic_dimensionless}) of the affine model, including quadrupolar self-gravity, becomes, in the incompressible limit,
\begin{equation}
  \f{\kappa^2}{\nu^2}=\tilde\rho x-(2\tilde\rho+1)x^2+\tilde\rho x^3,
\end{equation}
where we have used $H=Z/\sqrt{3}$, $\mathcal{W}=2/\sqrt{3}$ and $s=\tilde\rho/(\tilde\rho+1)$. This cubic function clearly agrees with the Taylor series for $f(x)$ up to the cubic term, but not beyond.

\bsp	
\label{lastpage}
\end{document}